\documentclass[aps,prb,twocolumn,amssymb]{revtex4}

\usepackage{graphicx}% Include figure files
\usepackage{dcolumn}% Align table columns on decimal point
\usepackage{bm}% bold math
\def\be{\begin{equation}}
\def\en{\end{equation}}
\def\bea{\begin{eqnarray}}
\def\ena{\end{eqnarray}}
\def\bec{\begin{equation}\begin{array}{rcl}}
\def\p{\partial}

\def\gs{\gtrsim}
\def\ls{\lesssim}
\def\ve{\varepsilon}
\def\bphi{\bar{\phi}}
\newcommand{\av}[1]{\langle{#1}\rangle}

\newcommand{\bi}[1]{\mbox{\boldmath$#1$}}

\renewcommand{\theequation}{\arabic{section}.\arabic{equation}}

\begin{document}
\title{ Phase Transitions in Soft Matter 
Induced by Selective Solvation
\footnote{Accounts: Bull. Chem. Soc. Jpn. June (2011)}}
\author{Akira Onuki,   Ryuichi Okamoto, and Takeaki Araki}
%\email[]{onuki@scphys.kyoto-u.ac.jp}
%\thanks{}
\address{Department of Physics, Kyoto University, Kyoto 606-8502,
Japan}


\date{\today}

\begin{abstract}
We  review our recent 
studies on selective solvation effects 
in phase separation  in polar binary mixtures 
with a small amount of solutes. 
Such hydrophilic or hydrophobic particles 
are preferentially attracted to one of 
the solvent components. 
We  discuss 
the role of  antagonistic salt composed of 
hydrophilic and hydrophobic ions, which undergo 
microphase separation at water-oil interfaces 
leading to mesophases. 
We then discuss phase separation induced by 
a strong selective solvent above 
a critical solute density $n_{\rm p}$, 
which occurs far from the 
solvent coexistence curve. 
We also give theories of  ionic surfactant systems 
and weakly ionized polyelectrolytes 
including  solvation among charged particles 
and polar molecules. We  point out  
 that the Gibbs formula 
 for the surface tension needs to 
include an electrostatic contribution 
in the presence of an 
electric double layer. 
\end{abstract}

\pacs{}
%\keywords{}

\maketitle


\section{Introduction}
\setcounter{equation}{0}

In  soft matter physics, much attention has been paid 
to the consequences of  the   
 Coulombic interaction among charged objects, 
such as  small ions,  charged  colloids, charged gels, 
and polyelectrolytes   
 \cite{Levin,Barrat,Holm,Rubinstein,PG,Safran}. 
However, not enough effort has  been made on 
 solvation effects among solutes 
(including hydrophobic particles) and polar solvent molecules 
 \cite{Is,Marcus,Gut,Chandler}. 
Solvation is also called 
hydration for water and  for aqueous mixtures. 
In  mixtures  of a water-like fluid   and a  less polar 
fluid (including polymer solutions),  
the  solvation is  preferential or selective, 
depending on whether the solute is hydrophilic or hydrophobic. 
See Fig.1 for its illustration. The  typical solvation 
free energy  much exceeds  the thermal energy $k_BT$ 
 per solute particle.  Hence  selective solvation 
 should  strongly influence   phase behavior 
 or even induce a new phase transition. 
In experiments on aqueous mixtures, it is well known that 
 a small amount of   salt  drastically alters  phase behavior 
 \cite{polar1,polar3,Misawa,Kumar,cluster,Taka,Anisimov}.  
 In biology, preferential  interactions 
 between water and cosolvents with proteins 
 are of crutial importance \cite{Tima,Tima1}. 
 Thus  selective solvation is relevant in  diverse 
 fields, but its  understanding  from physics is still 
in  its infancy.  
%With addition of a 
%10$^{-3}$ mole fraction of  NaCl, 
% the coexistence curve  is  typically shifted 
% by a few Kelvins  
%From our viewpoint,  the origins of 
%many mysterious  
%experimental findings  so far could be ascribed to 
%strong  selective solvation 
%in aqueous mixtures and 
%in polymeric systems containing a water-like solvent. 

%1 
\begin{figure}[htbp]
\vspace{-10mm}
\begin{center}
 \includegraphics[scale=0.32, bb= 0 0 1013 535]{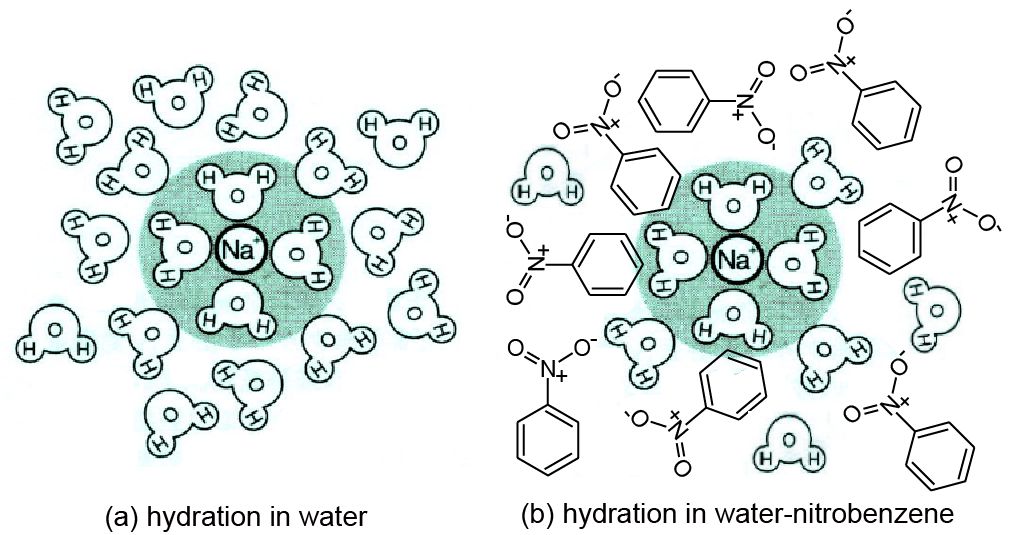}  
 \caption{Illustration of 
hydration of Na$^+$  surrounded 
by a shell composed of water molecules in 
(a) pure water and (b) water-nitrobenzene.  
 The solvation   chemical potential of Na$^+$  
 is higher  for (b) than for (a).}
\end{center}
\end{figure}

Around 1980, 
Nabutovskii {\it et al.}\cite{R1,R2} 
proposed a possibility of 
mesophases  in electrolytes 
from    a coupling between the composition and 
the  charge density in the free energy. 
In   aqueous mixtures, 
such a coupling originates 
from  the   selective solvation\cite{OnukiJCP04,OnukiPRE}. 
It is in many cases 
very strong, as suggested by  data of the Gibbs transfer 
free energy in electrochemistry (see Sec.2). 
Recently, several theoretical groups  have 
proposed Ginzburg-Landau  theories on   the  solvation  
in mixture solvents   for  electrolytes 
\cite{OnukiJCP04,OnukiPRE,OnukiJCP,Tsori,Ro1,Ro2,An1,Araki,Nara,Daan,Okamoto}, 
polyelectrolytes\cite{Onuki-Okamoto,Oka}, 
and ionic surfactants \cite{OnukiEPL}. 
In soft matter physics,  such coarse-grained  approaches   
 have been  used  to understand   cooperative  
 effects on mesoscopic scales \cite{PG,Safran,Onukibook}, 
 though  they are inaccurate on the angstrom scale. 
 They are even more usuful when 
     selective 
solvation comes into play in the strong coupling limit. 
This  review  presents 
such examples found in our recent research.

{\bf{1.1 Antagonistic salt}}.  An antagonistic salt  
consists  of hydrophilic and hydrophobic ions. 
An example is  
sodium tetraphenylborate  NaBPh$_{4}$, which   
dissociates into hydrophilic Na$^+$ and 
hydrophobic  BPh$_{4}^-$.  
The latter ion  consists  of  four phenyl rings bonded 
to an ionized boron. 
Such ion pairs in aqueous mixtures  behave 
antagonistically in the presence of composition 
heterogeneity. 
(i) Around a  water-oil interface,  they undergo 
microphase separation  on the scale of the Debye 
 screening length $\kappa^{-1}$, while   homogeneity holds  
 far from the interface to satisfy  
  the  charge neutrality (see the right bottom plate in Fig.2). 
This unique ion distribution produces    a large electric 
double layer and a large Galvani 
potential difference 
\cite{OnukiPRE,OnukiJCP,Araki,Nara}. 
We found that this ion distribution serves   to 
 much  decrease  the surface tension \cite{OnukiJCP}, 
in agreement  with experiments \cite{Reid,Luo}. 
From  x-ray   reflectivity measurements,  
Luo {\it et al.} \cite{Luo} determined 
 such ion distributions   
around a water-nitrobenzene(NB) interface 
by adding BPh$_{4}^-$ and two species of 
hydrophilic ions. (ii) In the vicinity of the solvent criticality, 
  antagonistic ion pairs 
 interact differently with water-rich and oil-rich 
composition fluctuations, leading to 
mesophases (charge density waves).   
In accord with  this prediction,  Sadakane {\it et al.} 
\cite{Sadakane,Seto} added a small amount of 
NaBPh$_{4}$     to a  near-critical 
mixture of D$_2$O and 3-methylpyridine (3MP) to find 
  a  peak   at an intermediate wave number 
$q_m$($\sim 0.1~$\AA$^{-1}\sim \kappa$) 
in the intensity  of small-angle neutron scattering. 
The peak height  was much 
enhanced with formation of periodic structures.    
(iii) Moreover, Sadakane {\it et al.}  observed  multi-lamellar (onion) 
structures   at  small volume fractions of 3MP 
(in D$_2$O-rich solvent)  far from 
the criticality \cite{SadakanePRL}, 
where BPh$_{4}^-$  and solvating 3MP form charged lamellae. 
These  findings  demonstrate 
  very strong hydrophobicity of BPh$_{4}^-$. 
(iv) Another interesting phenomenon 
is    spontaneous emulsification 
(formation of small water droplets) 
at a water-NB interface \cite{Aoki,Poland}.
It was observed when a large pure water droplet was pushed  into 
a cell  containing NB and  antagonistic salt 
(tetraalkylammonium chloride). This 
 instability was caused by 
ion transport through the interface.

{\bf{1.2 Precipitation due to selective solvation}}. 
Many experimental  
groups have  detected large-scale,  
 long-lived heterogeneities 
  (aggregates or domains)   
emerging with  addition of a hydrophilic salt or 
a hydrophobic solute in one-phase states of 
aqueous mixtures\cite{S1,S2,S3,S4,S5,S6,S7}.  
Their  small diffusion constants indicate that  
their typical size is  of order 
$10^3 {\rm \AA}$ at very small volume fractions. 
In two-phase states, they also observed a  third phase 
visible  as a thin solid-like 
plate  at  a liquid-liquid interface 
in two-phase states \cite{third}.
In our recent theory \cite{Okamoto}, 
for sufficiently  strong  solvation preference,  
a selective   solute 
can   induce  formation 
 of   domains rich in the selected component 
even very far  from  the solvent coexistence curve.  
This phenomenon occurs  when the 
volume fraction of the 
selected  component is relatively small. 
If it is a majority component,   
its aggregation is not needed. 
This precipitation phenomenon 
 should be widely  observable 
for various combinations of  solutes  
and mixture solvents.

{\bf{1.3 Selective hydrogen bonding}}. 
Hydrogen bonding is  
   of primary importance  in the 
phase behavior of soft matter. 
In particular, using statistical-mechanical theories, 
the origin of 
 closed-loop coexistence 
curves was  ascribed to the hydrogen bonding 
 for  liquid mixtures \cite{Wheeler,Goldstein} 
 and for  polymer 
solutions \cite{hydrogen1,hydrogen2}. Interestingly, 
water itself can be a selective solute 
triggering phase separation when  
 the hydrogen bonding differs significantly between the two 
components,  as observed in a mixture of 
methanol-cyclohexane \cite{Jacobs,Be}.     
More drastically, even 
 water absorbed from air  changed  
the phase behavior  
in films of  polystyrene(PS)- 
polyvinylmethylether(PVME) \cite{Hashimoto}.  
That is, a small amount of water 
induces precipitation of PVME-rich domains. 
For  block copolymers, 
similar precipitation of micelles can well be expected 
 when a small amount of water is added.

{\bf{1.4 Ionic surfactant}}. 
Surfactant molecules  are strongly  
trapped at an interface due to the amphiphilic 
interaction  even if 
their bulk density is very low \cite{PG-Taupin,Safran}. 
They can thus efficiently    
reduce  the surface tension,  giving rise to 
various mesoscopic 
structures. However, most theoretical studies 
have treated nonionic surfactants, 
while ionic surfactants  are important 
 in biology and technology. 
In this review, we also discuss  selective 
solvation in systems of  
ionic surfactants, counterions, and added ions 
in water-oil\cite{OnukiEPL}.   
 We shall see that the 
  adsorption behavior 
  strongly depends on the selective solvation.

{\bf{1.5 Polyelectrolytes}}. 
Polyelectrolytes are already very  complex 
because of  the electrostatic interaction  
among  charged particles (ionized monomers 
 and  mobile  ions)  
  \cite{Barrat,Holm,Rubinstein}.  Furthermore, 
we should take into account two ingredients\cite{Onuki-Okamoto}, 
which have not yet 
attracted enough attention. 
First,   the dissociation (or ionization) on the chains 
should be treated as a chemical 
reaction in many polyelectrolytes 
containing weak acidic  monomers \cite{Joanny,Bu1,Bu2}.   
Then the degree of ionization 
is a space-dependent annealed variable.  
Second, the solvation effects should come into play 
because  the  solvent molecules and 
 the  charged particles interact 
 via  ion-dipole  interaction. 
 Many  polymers themselves are   hydrophobic and  become 
 hydrophilic with progress of ionization in water. 
 This is because 
 the decrease of the free energy upon ionization 
 is very large. 
It is also worth noting that the selective solvation  
effect can be  dramatic  
in  mixture solvents\cite{Oka}.  
As an  example, precipitation of DNA has been  observed 
with addition of  ethanol in water 
\cite{B1,B2,B3}, where  the ethanol  added  is 
excluded from condensed DNA,   suggesting  
solvation-induced wetting of DNA by water. 
In polyelectrolyte solutions,  
macroscopic phase separation 
and mesophase formation  can both 
take place, sensitively depending on 
many parameters.

The organization of this paper is as follows. 
In Sec.2,   we will present   the  background 
of the solvation on the basis of some  experiments.    
In Sec.3,  we will explain 
a   Ginzburg-Landau model 
for electrolytes   
accounting for  selective solvation.  
In Sec.4, we will treat ionic surfactants 
by introducing  the amphiphilic interaction 
together with the solvation interaction. 
In Sec.5, we  will   examine precipitation 
induced by a strong selective solute, where 
  a simulation of the 
precipitation dynamics will also be presented. 
In Sec.6, we will give a Ginzburg-Landau model 
for weakly ionized polyelectrolytes 
accounting for  the ionization 
fluctuations and the solvation interaction. 
In  Appendix A, we will give a statistical theory of 
hydrophilic solvation at small water contents in oil.

\section{Background of selective solvation of ions}
{\bf {2.1 Hydrophilic ions in aqueous mixtures}}. 
Several  water  molecules  form  a solvation shell 
 surrounding a small  ion 
  via ion-dipole  interaction \cite{Is}, as in Fig.1.  
Cluster structures produced by solvation 
have been  observed by 
 mass spectrometric analysis \cite{W1,W2}.  
Here we mention an experiment by 
Osakai {\it et al}   \cite{Osakai}, which   
demonstrated   the presence of a solvation 
shell in a water-NB  mixture in two-phase coexistence. 
They  measured   the amount of water  
  extracted together with hydrophilic ions  in 
a NB-rich region   coexisting  with a salted water-rich  
region. A   water content of $0.168$M (the water 
volume fraction $\phi$ being 0.003)  
was already present in the NB-rich  phase  without ions. 
They  estimated   the average number of solvating water 
molecules in the  NB-rich phase 
to be  4 for Na$^+$, 
6 for Li$^+$, and 15 for Ca$^{2+}$ per ion. Thus,   
 when a hydrophilic  ion   moves from a  water-rich 
region to an  oil-rich region across an interface,  
a considerable fraction of 
 water molecules solvating it 
  remain attached to it. 
Furthermore, using proton NMR spectroscopy,   
Osakai {\it et al.} \cite{proton} 
studied successive formation 
of complex structures of anions X$^-$ (such as Cl$^-$ and Br$^-$) 
and water molecules 
by gradually increasing the water content in NB. 
This hydration  reaction is schematically 
written as  X$^-$(H$_2$O)$_{m-1}$ + H$_2$O 
 $\leftarrow\rightarrow$ 
X$^-$(H$_2$O)$_m$  ($m= 1, 2, 3, ...$). 
For  Br$^-$, these clusters    are 
 appreciable for   water content  larger 
than $0.1$M or for  water volume fraction $\phi$ exceeding 
 $0.002$.

In  Appendix A,  
we will calculate the  statistical distribution  of 
clusters composed 
of ions and polar molecules. 
Let  the  free energy typically decrease by 
$\epsilon_{{\rm b}i}$ upon binding 
 of a polar  molecule to a hydrophilic  ion of  species $i$. 
  A well-defined 
  solvation  shell is formed for $\epsilon_{{\rm b}i} \gg 
  k_BT$, where  the water volume fraction 
$\phi$ needs to satisfy   
\be 
\phi> \phi_{{\rm sol}}^i \sim  \exp(-\epsilon_{\rm{b}i}/k_BT). 
\en 
For $\phi\ll  \phi_{{\rm sol}}^i$ 
there is almost no solvation. 
%See  the appendix for a derivation of the above result, 
%where the  statistical distribution  of 
%clusters composed of ions and water molecules will be given.
The crossover volume fraction  
$ \phi_{{\rm sol}}^i$ is very small for strongly 
hydrophilic ions with $\epsilon_{\rm{b}i}\gg k_BT$. 
For  Br$^-$ in water-NB, we estimate 
$ \phi_{{\rm sol}}^i \sim 0.002$ from 
the experiment by Osakai {\it et al.} 
as discussed above\cite{proton}.

{\bf {2.2 Hydrophobic particles}}. Hydrophobic  objects 
are ubiquitous in nature, which 
repel  water  because of the 
strong attraction among hydrogen-bonded 
water molecules themselves \cite{Is,Chandler}. 
Hydrophobic particles  tend to form 
aggregates in water \cite{Pratt-hyd,Wolde-Chandler} and are 
more soluble in oil than in water.  
They  can be either neutral or charged. 
A widely used  hydrophobic anion  
is  BPh$_{4}^-$.    
In water, a large  hydrophobic particle $(\gs 1$nm)  
is even in  a cavity 
 separating the particle surface  from 
 water   \cite{Chandler}. 
In  a water-oil mixture,  on the other hand, 
hydrophobic particles 
 should be in contact with 
  oil molecules instead. This   attraction 
   can produce significant 
  composition heterogeneities on   
  mesoscopic scales around  hydrophobic objects,  
 which indeed  takes place  around protein surfaces 
  \cite{Tima,Tima1}.   
%  sometimes 
% leading to wetting transitions. 

{\bf {2.3 Solvation chemical potential}}.
We  introduce a solvation chemical potential 
$\mu_{\rm sol}^i(\phi)$  
 in the dilute limit of 
  solute  species $i$. It is the solvation part of the 
chemical potential of one particle (see Eq.(B1) 
in Appendix B). It is  a statistical average over the thermal 
fluctuations of the molecular configurations. 
In  mixture solvents,  it depends on the ambient 
water volume fraction  $\phi$. 
 For  planar  surfaces 
 or large particles (such as proteins),  
 we may consider the solvation 
 free energy per unit area.

 Born   \cite{Born} calculated   
the polarization energy of a polar fluid  
 around a hydrophilic ion  with charge $Z_ie$ 
  using   continuum  electrostatics  
%It follows from 
% the space integral of the electrostatic energy density 
%$\ve {\bi E}^2/8\pi \propto r^{-4}$, 
%where ${\bi E}=-(Z_i e/\ve  r^{3}) {\bi r}$  
%is the electric field around the ion. 
to obtain the classic   formula,    
\be  
(\mu_{\rm sol}^i)_{\rm Born} 
=  -({Z_i^2e^2}/{2R_{i}}) (1-1/\ve).  
\en 
The contribution without polarization ($\ve=1$)  or in 
vacuum is subtracted    
%The right hand side 
%considerably exceeds $k_BT$ even for 
%$\ve = 80$ at room temperatures 
and the $\phi$ dependence here arises from that of   
the  dielectric constant $\ve= \ve(\phi)$  
(see Eq.(3.5) below).  
The lower cutoff    $R_{i}$ 
 is   called the Born radius, which is  on the  order of 
 $1{\rm \AA}$ for small metallic ions  \cite{Is,RMarcus}. 
The hydrophilic solvation   is stronger 
for smaller ions, since it arises from 
the ion-dipole interaction. 
In this original formula,  neglected are  
 the formation of a solvation shell, the density and 
 composition changes (electrostriction), and 
 the nonlinear dielectric effect.

For mixture solvents, the  binding free energy 
between a hydrophilic ion and a polar molecule is 
estimated from the  Born formula (2.1) as 
 $\epsilon_{\rm{b}i} \sim - 
\p [(\mu_{\rm sol}^i)_{\rm Born}]/\p\phi$ or 
\cite{OnukiJCP04}
\be 
\epsilon_{\rm{b}i} \sim Z_i^2e^2\ve_1/R_{i}\ve^2 
= k_{B}T Z_i^2\ell_B\ve_1/R_{i}\ve,
\en 
where  $\ve_1=\p \ve/\p \phi$ 
and $\ell_B=e^2/k_BT\ve$ 
is the Bjerrum length 
($\sim 7{\rm \AA}$ for water at room temperatures). 
For  $\ve_1\sim \ve$, a well-defined shell 
appears  for $Z_i^2\ell_B \gg R_i$. 
The solvation chemical 
potential $\mu_{\rm sol}^i(\phi)$ 
of   hydrophilic ions  in water-oil 
should largely decrease to negative values 
in the narrow  range 
$0<\phi< \phi_{\rm sol}^i$,  as will be 
described in Appendix A.
In the wide composition range (2.1), 
its  composition dependence  is  still very 
strong such that $|\p \mu_{\rm sol}^i/\p \phi| 
\gg k_BT$ holds (see the next subsection). 
The solubility of hydrophilic  ions  should also 
increase abruptly in the narrow range $\phi< \phi_{\rm sol}^i$ 
with addition of water to oil. Therefore, 
solubility measurements of hydrophilic ions 
would be informative 
at very small water contents in oil.

In sharp contrast,  $\mu_{\rm sol}^i$ 
of a neutral hydrophobic particle  
 increases with increasing the particle  radius $R$ 
   in  water \cite{Chandler}. It 
  is roughly proportional to 
the surface area  $4\pi R^2$ for 
$R \gs  1$nm and is  estimated to be 
 about  $100k_BT$ at $R \sim 1$nm. 
In water-oil solvents, on the other hand, it is not easy to 
estimate the $\phi$ dependence of 
$\mu_{\rm sol}^i$. 
However,  $\mu_{\rm sol}^i$ 
  should strongly increase with 
increasing the water composition $\phi$, 
since  hydrophobic particles (including ions) 
 effectively attract 
 oil molecules\cite{Tima1,SadakanePRL}.

{\bf{2.4 Gibbs transfer free energy}}. 
We consider  a liquid-liquid  interface  between  
a  polar (water-rich) 
phase $\alpha$ and a less polar (oil-rich) 
phase $\beta$ with bulk compositions $\phi_\alpha$ 
and $\phi_\beta$ with $\phi_\alpha>\phi_\beta$.  
The solvation chemical potential  
$\mu_{\rm sol}^i(\phi)$ takes  different values 
in the two phases due to its 
 composition dependence. So we define    
\be 
\Delta\mu_{\alpha\beta}^{i}
= \mu_{\rm sol}^{i}(\phi_\beta)- 
\mu_{\rm sol}^{i}(\phi_\alpha).  
\en 
In electrochemistry \cite{Hung,Ham,Koryta,Osakai,Sabela},  
the difference of the solvation free energies  
 between two coexisting phases 
is  called the standard 
Gibbs transfer free  energy  denoted by 
   $\Delta G_{\alpha\beta}^{i}$    for each  
  ion species $i$.  
 Since it  is  usually measured in units of 
kJ per mole,  its division  
by the Avogadro number $N_A$ gives  
 our  $\Delta\mu_{\alpha\beta}^{i}$ or    
$\Delta\mu_{\alpha\beta}^{i}=\Delta G_{\alpha\beta}^{i}/N_A$. 
With $\alpha$ being the water-rich phase, 
$\Delta \mu_{\alpha\beta}^i$ is positive  
for hydrophilic ions and is negative 
for hydrophobic  ions from its definition (2.4). 
See Appendix B for  relations between  
$\Delta\mu_{\alpha\beta}^{i}$ 
and other  interface quantities.

For ions, most data of  
$\Delta G_{\alpha\beta}^i$ 
are at present  limited on 
water-NB 
\cite{Hung,Koryta,Osakai}  and water-1,2-dichloroethane(EDC) 
\cite{Sabela}  at room temperatures, where 
the dielectric constant of NB ($\sim 35$) 
is larger than that of  EDC ($\sim  10$).  They then yield the ratio  
  $\Delta\mu_{\alpha\beta}^i/k_{B}T$. 
In the case of water-NB,  
it is $13.6$  for Na$^+$,   $27.1$  for Ca$^{2+}$, and 
$11.3$ for Br$^-$  as examples of 
hydrophilic ions, while it is $-14.4$ for  hydrophobic  
BPh$_4^-$. In the case of  water-EDC, 
it is  $22,7$  for Na$^+$ and   $17.5$  for Br$^-$, 
while it is $-14.1$ for    BPh$_4^-$.  
The amplitude  $|\Delta\mu_{\alpha\beta}^i|/k_{B}T$ for 
hydrophilic ions 
 is larger  for EDC than for NB and is very large for 
 multivalent ions.   
Interestingly,  $\Delta\mu_{\alpha\beta}^i$ 
 for  H$^+$ (more precisely  hydronium ions H$_3$O$^+$) 
 assumes  positive values close to those for  Na$^+$ 
 in  these two mixtures.

In these experiments on ions,  a hydration shell  
should have been formed around hydrophilic ions  
even in the water-poor phase $\beta$ (presumably not completely\cite{proton}). 
From Eq.(2.1)  $\phi_\beta$ should be  exceeding  the crossover 
volume fraction $\phi_{\rm{sol}}^i$. 
The data  of 
the Gibbs transfer free energy quantitatively 
demonstrate very strong selective solvation 
  even in the range (2.1) or after the  shell formation.

On the other hand,  neutral hydrophobic particles 
are less soluble in a water-rich phase  $\alpha$ than in 
an oil-rich phase $\beta$. Their  chemical potential 
is given by 
$
\mu_i= k_BT \ln (n_i\lambda_i^3) 
+ \mu_{\rm sol}^i (\phi),
$ 
where $i$ represents the particle species and 
$\lambda_i$ is the thermal 
de Broglie wavelength. From homogeneity of $\mu_i$ 
across an interface, we obtain the ratio of their 
equilibrium bulk 
 densities  as  
\be 
n_{i\beta}/n_{i\alpha}= \exp[- \Delta\mu_{\alpha\beta}^{i}/k_BT], 
\en 
where $\Delta\mu_{\alpha\beta}^i<0$ from the definition (2.4).

\setcounter{equation}{0}
\section{Ginzburg-Landau theory of mixture electrolytes}

{\bf {3.1 Electrostatic and solvation interactions}}. 
We present a  Ginzburg-Landau 
free energy $F$ for  a polar binary 
 mixture  (water-oil)  
containing a small amount of a monovalent 
 salt ($Z_1=1$, $Z_2=-1$). 
 The multivalent case should be studied separately.  
  The ions  are  dilute and 
 their  volume fractions are  negligible, 
 so we are allowed to neglect 
 the formation of ion  clusters \cite{Levin,pair1,pair2}.
The variables  $\phi$, $n_1$, and $n_2$ are coarse-grained 
ones varying  smoothly on the molecular scale. 
For simplicity, we also  neglect the image 
interaction \cite{Onsager,Levin-Flores}. 
%though it  was included in our previous papers 
%\cite{OnukiPRE,OnukiJCP}. 
At  a water-air interface 
the  image interaction 
% is relevant  
%at very low ion densities, 
 serves to push ions 
into the water region. However, 
hydrophilic ions are already 
strongly depleted from  an interface 
due to their 
position-dependent hydration 
 \cite{Levin-Flores}. See our previous analysis 
 \cite{OnukiPRE} 
 for  relative importance between the image interaction 
 and the solvation interactiion at a liquid-liquid interface.

The free energy $F$ is the space integral of 
 the free energy density of the form, 
\be
F =\int d{\bi r}\bigg[f_{\rm tot} + \frac{1}{2}C |\nabla\phi|^2 
+ \frac{\varepsilon }{8\pi }{\bi E}^2 \bigg].  
\en
The first term $f_{\rm tot}$ depends  on 
$\phi$, $n_1$, and $n_2$ as   
\be
{f_{\rm tot}} = f(\phi) 
 + k_B T\sum_i n_i  \bigg[\ln (n_i\lambda_i^3) -1-  g_i \phi\bigg].
\en 
In this paper, the solvent molecular volumes of 
the two components are assumed to take 
a common value $v_0$, though 
they are often  very different 
in real binary mixtures. 
%For example, those  of 
% D$_{2}$O and 3MP (the inverse densities of 
%the pure components) are 28 and 
% 168 \AA$^3$, respectively. 
Then $f $ is of   
the    Bragg-Williams form \cite{Safran,Onukibook},      
\be 
\frac{v_0f}{k_BT}   =   
 \phi \ln\phi + (1-\phi)\ln (1-\phi) 
+ \chi \phi (1-\phi),   
\en 
where   $\chi$ is the interaction 
parameter dependent on   $T$. 
The critical value of $\chi$ is 2 without ions. 
The $\lambda_i = \hbar(2\pi/m_i k_BT)^{1/2}$ in Eq.(3.2) 
is   the thermal de Broglie wavelength of the species $i$ 
with $m_i$ being the molecular mass. 
The $g_1$ and $g_2$ are the solvation coupling constants. 
In  addition, 
the coefficient $C$  in the  gradient part of Eq.(3.1) 
remains an arbitrary constant.  
To explain experiments, 
however, it is desirable to 
determine $C$  from the surface tension data 
or from the scattering data.

In the electrostatic part  of Eq.(3.1), 
the electric field is written as   ${\bi E}=-\nabla\Phi$. 
The  electric   potential $\Phi$ 
satisfies   the   Poisson equation,    
\be 
-\nabla\cdot\ve\nabla \Phi=  4\pi \rho. 
\en 
The dielectric constant $\ve$ is assumed to 
depend on $\phi$   as    
\be 
\ve(\phi)=\ve_0 + \ve_1 \phi. 
\en 
where $\ve_0$ and  $\ve_1$ are positive constants. 
Though there is no reliable theory of $\ve(\phi)$  
for a polar mixture,  
a linear composition dependence of $\ve(\phi)$ 
was  observed by Debye and Kleboth 
for a mixture of 
nitrobenzene-2,2,4-trimethylpentane \cite{Debye}. 
In addition, the form of the electrostatic part of the free energy 
density depends on the experimental method \cite{Tojo}. 
Our form in Eq.(3.1) follows 
if we insert the fluid between parallel plates 
 and fix the charge densities 
on the two plate surfaces.

We explain the   solvation terms  in $f_{\rm tot}$ 
in more detail. They   follow 
if   $\mu_{\rm sol}^i(\phi)$ ($i=1,2$) 
 depend on $\phi$ linearly as  
\be 
\mu_{\rm sol}^i(\phi) =A_i   -k_B Tg_i\phi.   
\en  
Here the first term    $A_i$  is a constant  yielding    
 a  contribution linear with respect to $n_i$ in $f_{\rm tot}$, so it  
 is irrelevant at constant ion numbers. The  second term 
 gives rise to  the solvation  coupling  in $f_{\rm tot}$. 
In this approximation, $g_i>0$ 
for hydrophilic ions and $g_i<0$ 
for hydrophobic ions.  
The difference of the solvation 
chemical potentials  in two-phase coexistence  in Eq.(2.4) 
 is given by  
\be 
\Delta\mu_{\rm sol}^i= k_B Tg_i\Delta\phi,
\en 
where $\Delta\phi=\phi_\alpha-\phi_\beta$ is the 
composition difference. From Eqs.(B4) and (B5), 
the Galvani potential difference  is 
\be 
\Delta\Phi= k_BT (g_1-g_2)\Delta\phi /2e,
\en 
and  the ion reduction factor is 
\be 
{n_{1\beta}}/{n_{1\alpha}}=
{n_{2\beta}}/{n_{2\alpha}}= 
\exp[- (g_1+g_2){\Delta\phi}/{2}].
\en   
The  discussion in 
 subsection 2.3 indicates   
 $g_i \sim 14$  (23) for Na$^+$ ions and 
 $g_i \sim -14$ (-14) 
 for  BPh$_4^-$ in water-NB (water-EDC)  at 300K. 
For multivalent ions 
$g_i$  can be very large 
($g_i \sim 27$  for Ca$^{2+}$ 
in water-NB). 
The linear form (3.6) 
is adopted for the mathematical 
simplicity and is valid for $\phi>\phi_{\rm sol}^i$ 
after the solvation shell formation (see Eq.(2.1)). 
The results in Appendix A 
suggest a more complicated 
functional form of $\mu_{\rm sol}^i(\phi)$.

In equilibrium,  we require the homogeneity of 
the chemical potentials $h= \delta F/\delta\phi$ and  
$\mu_i= \delta F/\delta n_i$. 
Here,  
\bea 
&&h=f' -C\nabla^2\phi -\frac{\ve_1}{8\pi}{\bi E}^2 
- k_BT \sum_i g_i n_i,\\
&&\mu_i= k_BT [\ln(n_i\lambda_i^3)- g_i\phi]+ Z_ie\Phi ,
\ena 
where $f'=\p f/\p \phi$. The  ion distributions are 
expressed in terms of  $\phi$ and $\Phi$ 
in the modified Poisson-Boltzmann relations \cite{OnukiPRE},  
\be 
n_i=  n_i^0 \exp[g_i \phi -Z_i e\Phi/k_BT].
\en
The coefficients 
 $n_i^0$ are  determined 
from the conservation of the ion numbers, 
$\av{n_i}= V^{-1} 
\int d{\bi r}n_i({\bi r})=n_0,
$  
where $\av{\cdots}=V^{-1}\int d{\bi r}(\cdots)$ 
denotes the space average 
with $V$ being the cell volume. The average $n_0
=\av{n_1}=\av{n_2}$ 
is a given constant density in the monovalent case.

It is worth noting that  
a similar Ginzburg-Landau free energy 
was proposed by Aerov {\it  et al}. \cite{Aerov}  
for mixtures  of ionic and nonionic liquids 
composed of anions, cations, and 
water-like molecules. 
In such mixtures,  the interactions 
among neutral molecules and ions 
can be preferential, leading to  
mesophase formation, as   
has been predicted also by  molecular dynamic simulations 
\cite{Ionic}.

{\bf {3.2  Structure factors  and effective 
ion-ion interaction  in one-phase states}}. 
The simplest application of our model is to 
calculate the structure factors of the 
the composition  and the ion densities in 
 one-phase states. They can be measured by scattering 
 experiments. 
 
We  superimpose  small deviations 
$\delta\phi({\bi r})= \phi({\bi r})-\av{\phi}$ 
and $\delta n_i({\bi r})= n_i({\bi r})-n_0$ 
on the averages $\av{\phi}$ and $n_0$. 
The monovalent case $(Z_1=1, Z_2=-1)$ is treated. 
As thermal fluctuations, the statistical  distributions 
of  $\delta\phi({\bi r})$ and $\delta n_i$ are Gaussian  
 in the mean field theory. 
We may neglect the composition-dependence of $\ve$ for such small 
deviations.  We calculate the following, 
\bea 
S(q)&=& {\av{|\phi_{\small{\bi q}}|^2}}_{\rm e},\quad 
G_{ij}(q)= {\av{n_{\small{i\bi q}} n_{j\small{\bi q}}^*   }}_{\rm e}/ n_0, 
\nonumber\\ 
C(q)&=&
{\av{|\rho_{\small{\bi q}}|^2}}_{\rm e}/e^2n_0,   
\ena 
where  $\phi_{\small{\bi q}}$, 
 $n_{i\small{\bi q}}$, and $\rho_{\small{\bi q}}$  
 are the Fourier components  of 
 $\delta\phi$,  $\delta n_i$, 
 and the charge density $\rho=e(n_1-n_2)$ 
with wave vector $\bi q$ and 
 $\av{\cdots}_{\rm e}$ 
denotes taking the thermal average. 
 We introduce the  Bjerrum length $\ell_B=  e^2/\ve k_B T$ 
and the Debye wave number $\kappa=(8\pi  \ell_Bn_0)^{1/2}$.

First, the inverse of $S(q)$ is written as\cite{OnukiJCP04,OnukiPRE}   
\be
\frac{1}{S(q)}
={\bar r}- (g_1+g_2)^2\frac{n_0}{2}  + \frac{Cq^2}{k_BT}  
\bigg[1-  \frac{\gamma_{\rm p}^2\kappa^2}{q^2+\kappa^2}\bigg] , 
\en
where ${\bar r}= f''/k_BT$ with 
$f''= \p^2 f/\p \phi^2$. The second term 
is large for large $(g_1+g_2)^2$ even for small 
average ion density $n_0$, giving rise to a large 
shift of the spinodal curve.  If $g_1 \sim g_2 \sim 15$, 
this factor is of order $10^3$.  
In the previous experiments
 \cite{polar1,polar3,Misawa,Kumar,cluster,Taka,Anisimov}, 
 the shift of the coexistence curve is 
typically a few Kelvins with addition of a 
10$^{-3}$ mole fraction of a hydrophilic salt like NaCl.  
The parameter $\gamma_{\rm p}$ in the third term 
represents asymmetry of the solvation 
of the two ion species and is defined by 
\be
\gamma_{\rm p}= (k_BT/16 \pi C \ell_{B})^{1/2} 
|g_1-g_2|. 
\en  
If the right hand side of Eq.(3.14) is expanded  
with respect to $q^2$, the coefficient of 
$q^2$ is $C(1-\gamma_{\rm p}^2)/k_BT$. 
Thus  a Lifshitz point is realized at $\gamma_{\rm p}=1$. 
For $\gamma_{\rm p}>1$,  $S(q)$ has a peak at 
an intermediate wave number,
\be 
q_{\rm m}=( \gamma_{\rm p}-1)^{1/2}\kappa. 
\en  
The peak height $S(q_{\rm m})$ 
and the long wavelength limit $S(0)$ are related by 
\be 
1/S(q_{\rm m})=1/S(0) - C ( \gamma_{\rm p}-1)^{2} \kappa^2/k_BT.
\en  
A mesophase appears with decreasing $\bar r$ or increasing $\chi$, 
 as observed by Sadakane {\it et al.} \cite{Sadakane}. 
 In our mean-field theory, 
  the criticality  of a binary mixture 
 disappears if  a salt with   
$\gamma_{\rm p}>1$ is added however small its content is.  
%It is worth noting that our  
%$S(q)$  is analogous to 
% that for  weakly ionized 
%polyelectrolytes, for which see Sec.6. 
Very close to the solvent criticality, 
Sadakane {\it et al.} \cite{Seto}. 
recently  measured   anomalous  scattering 
from D$_2$O-3MP-NaBPh$_{4}$ 
stronger than that from D$_2$O-3MP 
without NaBPh$_{4}$. There, 
 the observed scattering  amplitude  is not  well described 
by our $S(q)$ in Eq.(3.14), requiring 
more improvement.

Second, retaining   the fluctuations of the 
ion densities, we  eliminate 
 the composition  fluctuations in $F$ 
to obtain  the effective  interactions 
among  the ions mediated by the composition fluctuations. 
The resultant free energy of ions is written as \cite{OnukiPRE}   
\bea 
&&{ F}_{\rm ion} = 
\int d{\bi r} 
  \sum_{i}  {k_BT}n_i \ln (n_i\lambda_i^3)  \nonumber\\
&&\hspace{-1cm} +  \frac{1}{2} 
\int\hspace{-1mm} d{\bi r}\hspace{-1mm}\int\hspace{-1mm} d{\bi r}' 
 \sum_{i,j}  V_{ij}(|{\bi r}-{\bi r}'|) 
 \delta n_{i}({\bi r})\delta n_{j}({\bi r}' ) .   
\ena 
The effective interaction 
potentials ${V_{ij}(r)}$ are given by 
\be 
{V_{ij}(r)}= Z_iZ_j 
\frac{e^2}{\ve r}-\frac{g_ig_j }{A}
\frac{1}{r}{e^{-r/\xi}} ,
\en 
where $Z_1$ and $Z_2$ are  $\pm 1$ in the monovalent case,  
$A= 4\pi C/(k_BT)^2$, and  
$\xi= (C/{\bar r})^{1/2}$ is 
the correlation length.  
The second term in Eq.(3.19) arises  from 
the selective solvation and  
is effective in the range $a\ls r\ls \xi$ 
and can be increasingly 
important on approaching 
the solvent criticality (for $\xi\gg a$).  It 
is attractive among the ions of the same species $(i=j)$ 
 dominating  over the  Coulomb 
repulsion  for 
\be 
g_i^2> 4\pi C \ell_B/k_BT.
\en 
Under the above  condition   
there should be a  tendency of ion  aggregation 
of the same species.  In the antagonistic case 
($g_1g_2<0$), the cations and anions 
can repel one another in the range $a\ls r\ls \xi$  for
\be 
|g_1g_2|> 4\pi C \ell_B/k_BT,
\en  
under which  charge density waves are triggered near 
the solvent criticality.

The ionic structure factors 
can readily be calculated from Eqs.(3.18) and (3.19). 
Some calculations give  \cite{Nara} 
\bea 
&&\hspace{-3mm}\frac{G_{ii}(q)}{G_0(q)}=1 + {n_0}S(q)\bigg[ 
g_1^2+g_2^2- \frac{(g_1-g_2)^2}{2(u+1)}-
 \frac{2g_i^2u}{2u+1} \bigg],\nonumber\\
&&\hspace{-3mm}{G_{12}(q)}=\frac{1}{2} + \frac{n_0}{4}S(q) 
(g_1+g_2)^2- \frac{1}{4}C(q),  \nonumber\\
%&&{G_{12}(q)}=\frac{1}{2(u+1)} + \frac{n_0}{4}S(q)\bigg[ 
%(g_1+g_2)^2- \frac{(g_1-g_2)^2u^2}{(u+1)^2} \bigg],  
&&\hspace{-3mm}{C(q)}= 
\frac{2u}{u+1}+ {n_0}S(q)\frac{(g_1-g_2)^2}{(u+1)^2}u^2,
\ena  
where $u=q^2/\kappa^2$ and 
\be 
G_0(q)= \frac{u+1/2}{u+1} = 
\frac{q^2+\kappa^2/2}{q^2+\kappa^2}
\en 
is the structure factor for the cations (or 
for the anions) divided by $n_0$  in the absence of 
solvation.  The solvation parts in Eq.(3.22)  
are all proportional to $n_0S(q)$, where $S(q)$ 
is given by Eq.(3.14). The Coulomb 
interaction suppresses 
large-scale charge-density fluctuations, 
so $C(q)$ tends to zero as $q\to 0$.

It should be noted  that de Gennes \cite{deGennes} 
derived  the effective interaction among monomers 
on a chain ($\propto -e^{-r/\xi}/r$) mediated by 
the composition fluctuations in a mixture solvent. 
He then predicted 
anomalous size behavior of a chain  near the solvent 
criticality. However, as shown in Eq.(3.19), 
the  effective interaction is much more 
amplified  among charged particles 
than among neutral  particles. This indicates 
importance of the selective solvation for  a charged polymer  
in  a mixture solvent, even leading to a prewetting transition 
around a chain \cite{Oka}. 
We also point out that 
an attractive interaction arises among charged 
colloid particles due to 
the selective solvation 
in a mixture solvent, 
on which we will report shortly.

{\bf {3.3 Liquid-liquid interface profiles}}. 
The second application is to 
calculate a one-dimensional  liquid-liquid interface   
at $z=z_0$ taking the  $z$ axis in 
 its normal  direction, where   
$\phi \to \phi_\alpha$ in water-rich phase $\alpha$ 
($z-z_0 \to -\infty$) and $\phi \to \phi_\beta$ 
in oil-rich phase $\beta$ ($z-z_0 \to \infty$). 
In Fig.2, we give numerical results of typical  interface profiles,  
where we measure space  
in units of $a\equiv v_0^{1/3}$ 
and set $C=2.5 k_BT/a^2$,  
$e^2/\ve_0k_B T= 3a$,  and $\ve_1=\ve_0$.  
In these  examples,  the 
correlation length $\xi$ 
 is shorter than the Debye lengths  
$\kappa_\alpha^{-1}$ and $\kappa_\beta^{-1}$ in the two phases. 
However,  near the solvent criticality,  
$\xi$ grows above $\kappa_\alpha^{-1}$ and $\kappa_\beta^{-1}$ 
and we encounter another regime, 
which is not treated in this review.

%2
\begin{figure}[htbp]
\vspace{-30mm}
\begin{center}
\includegraphics[scale=0.45, bb= 0 0 696 714]{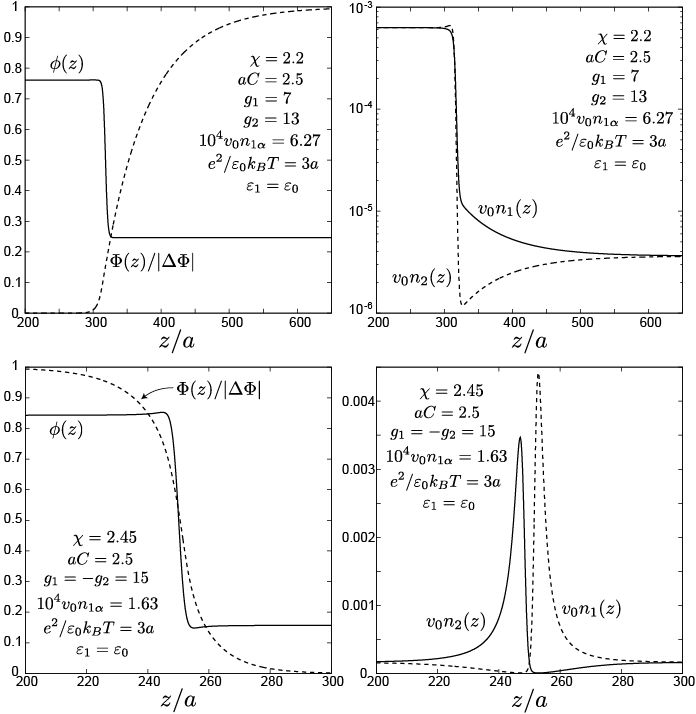}
\caption{ 
Top: normalized potential 
$\Phi(z)/\Delta\Phi$ (left),  composition $\phi(z)$ 
(left), and normalized ion densities 
$v_0 n_1(z)$  and $v_0n_2(z)$ 
(right on a semilogarithmic scale) for hydrophilic 
ion pairs,  where  $\chi =2.2, g_1=7, g_2=13$, 
 $v_0n_{1\alpha} =v_0n_{1\beta}
=4\times 10^{-4}$, and $e\Delta \Phi=-1.55 k_BT$. 
Bottom: those  for  antagonistic ion pairs, 
 where  $\chi =2.45, g_1=15, g_2=-15$, 
 $v_0n_{1\alpha} =v_0n_{1\beta}
=1.63 \times 10^{-4}$, and $e\Delta \Phi=10.3 k_BT$. 
In these cases $e^2/\ve_0k_B T= 3a$ and $\ve_1=\ve_0$. 
}
\end{center}
\end{figure}

The  upper plates  give  
 $\phi(z)$, $\Phi(z)$, 
$n_1(z)$, and $n_2(z)$ for hydrophilic ion pairs 
with $g_1= 7$ and $g_2=13$ at $\chi=2.2$ 
and $n_{1\alpha}=n_{2\alpha}= 
4\times 10^{-4}v_0^{-1}$. The ion reduction factor 
in Eq.(3.9) is $0.0057$.  
The potential $\Phi$ varies 
 mostly in the right side (phase $\beta$)  
on the scale of the Debye  length 
$\kappa_\beta^{-1}=67.5a$ (which is much 
longer than that $\kappa_\alpha^{-1}=6.1a$ 
in phase $\alpha$). The Galvani potential 
difference  $\Delta \Phi$ is $0.76 k_BT/e$ 
 here. The surface tension 
here is $\sigma= 6.53\times  10^{-2}k_BT/a^2
$  and is slightly larger than 
that $\sigma_0= 6.22k_BT/a^2$ 
without ions (see the next subsection).

The lower plates 
 display the same quantities 
for antagonistic  ion pairs 
with $g_1= 15$ and $g_2=-15$ for  $\chi=2.45$  
and $n_{1\alpha}=n_{2\alpha}= 
1.67\times 10^{-4}v_0^{-1}$. The anions 
and the cations are undergoing 
 microphase separation at the 
interface on the scale of the Debye  
lengths   $\kappa_\alpha^{-1}=12.3a$ 
and $\kappa_\beta^{-1}=9.70a$,   resulting in a large    
electric double layer and   a large  
potential drop ($\sim 10k_BT/e$). The surface tension 
here is $\sigma=0.0805 k_BT/a^2$ and  is about  
half of that $\sigma_0= 0.155k_BT/a^2$ without ions. 
This large decrease in $\sigma$  is marked in view of small $n_{1\alpha}$. 
A large decrease of the surface tension was observed for 
an antagonistic salt \cite{Reid,Luo}

{\bf {3.4 Surface tension}}. 
There have been numerous measurements of   the 
surface tension of an air-water interface with a 
salt in the water region. 
In this case,  almost all salts 
lead to an increase in the 
surface tension \cite{Onsager,Levin-Flores}, 
while acids tend to lower 
it because hydronium ions 
are trapped 
at an air-water interface.\cite{h1,Levinhyd}

Here, we consider the surface tension 
of a liquid-liquid interface 
in our Ginzburg-Landau scheme, where ions can be 
present in the two sides of the interface. 
In equilibrium we  minimize  
 $\Omega=\int d{z}\omega$, 
where $\omega$  is the grand potential density, 
\be 
\omega= f_{\rm tot}  + \frac{1}{2} C |\nabla\phi|^2 
+ \frac{\varepsilon }{8\pi }{\bi E}^2  
- h\phi- \sum_i \mu_in_i.
\en 
Using  Eqs.(3.10) and (3.11) we 
find $d(\omega+\rho \Phi)/dz= 2C\phi'\phi''$, 
where $\phi'=d\phi/dz$ and $\phi''=d^2\phi/dz^2$.
Thus,   
\be 
\omega= C\phi'^2-\rho\Phi +\omega_\infty, 
\en  
Since $\phi'$ and $\rho$ tend to zero far from the interface, 
 $\omega(z)$ tends to 
a common constant $\omega_\infty$ as $z\to \pm\infty$. 
The surface tension $\sigma=\int dz [\omega(z)-\omega_\infty]$ 
is then written  as \cite{OnukiPRE,OnukiJCP}  
\be 
\sigma =  \int dz \bigg[ C\phi'^2- \frac{\ve }{4\pi}{\bi E}^2\bigg]
= 2\sigma_{\rm g} -2\sigma_{\rm e}, 
\en  
where we introduce  the  areal densities of the gradient free energy 
  and the electrostatic energy   as 
\be 
\sigma_{\rm g}= \int dz~ C\phi'^2/2, \quad 
\sigma_{\rm e}=   \int dz~ {\ve }{\bi E}^2/{8\pi}. 
\en 
The expression $\sigma= 2\sigma_{\rm g}$ is well-known in the 
Ginzburg-Landau theory without the  electrostatic 
interaction \cite{Onukibook}.

%3
\begin{figure}[htbp]
\vspace{-15mm}
\begin{center}
\includegraphics[scale=0.59, bb= 0 0 490 353]{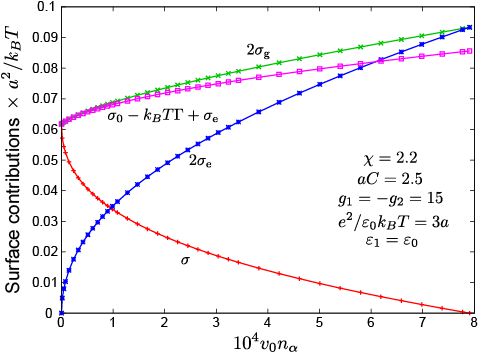}
\caption{Surface quantities $\sigma$, $2\sigma_{\rm e}$, $2\sigma_{\rm g}$, 
and $\sigma_0-k_BT\Gamma+ \sigma_e$ in units of $k_BTa^{-2}$ 
vs $v_0 n_\alpha$ for a monovalent, antagonistic 
 salt with $g_1=-g_2=15$ at $\chi=2.4$. Here $2\sigma_{\rm g}$ 
 and $\sigma_0-k_BT\Gamma+ \sigma_e$  are close,   
 supporting  Eq.(3.30).  Growth of $2\sigma_{\rm e}$ 
 gives rise to vanishing of the surface tension 
 $\sigma$ at $v_0  n_{\alpha}=8\times 10^{-4}$.  
 }
\end{center}
\end{figure}
 
In our previous work \cite{OnukiPRE,OnukiJCP}, 
we obtained the following approximate expression 
for  $\sigma$ valid for small ion  densities:  
\be 
\sigma\cong \sigma_0 -k_BT\Gamma - \sigma_{\rm e},
\en 
where $\sigma_0$ is the surface tension without 
ions and  $\Gamma$ is the  adsorption 
to the interface.  In terms of the total 
ion  density $n=n_1+n_2$, it may be expressed as  
\be 
\Gamma= \int dz \bigg[ n-n_\alpha -\frac{\Delta n}{\Delta\phi}
 (\phi-\phi_\alpha)\bigg],
\en 
where $n_K=n_{1K}+n_{2K}$ ($K=\alpha$, $\beta$), 
$\Delta n=n_\alpha-n_\beta$, and  the integrand  
tends to zero  as $z\to \pm\infty$. 
From   Eqs.(3.26) and (3.28)     
$\sigma_{\rm g}$ is expressed 
at small ion densities as  
\be 
2\sigma_{\rm g} \cong \sigma_0 -k_BT\Gamma +\sigma_{\rm e}. 
\en  
In the  Gibbs formula 
($\sigma\cong \sigma_0 -k_BT\Gamma$)  
\cite{Safran,Gibbs}, 
the electrostatic contribution  $-\sigma_{\rm e}$  is neglected. 
However, it   is  crucial 
 for antagonistic salt\cite{OnukiJCP,Araki}   and for 
ionic surfactant \cite{OnukiEPL}.

In Fig.3,  numerical results of  
$2\sigma_e$,  $2\sigma_{\rm g}$, 
$\sigma$, 
and the combination 
$\sigma_0 -k_BT\Gamma +\sigma_{\rm e}$ are plotted  
as functions of 
the bulk ion density $n_\alpha=n_{1\alpha}+n_{2\alpha}$ 
for the antagonistic case  $g_1=- g_2=15$, where 
$\chi=2.4$  and $C=2.5 k_BT /a^2$. 
The parameter $\gamma_{\rm p}$ 
in Eq.(3.15) exceeds unity (being equal to 1.89 
for $\ve= 1.5 \ve_0$). 
In this example,  
$\sigma_{\rm g}$  weakly depends  on $n_{\alpha}$ 
and  is fairly in accord with Eq.(3.30), 
while $\sigma_{\rm e}$ steeply increases 
with increasing $n_{\alpha}$. As a result, 
 $\sigma_{\rm e}$ increases up to  $\sigma_{\rm g}$,  
 leading to vanishing of  $\sigma$   
at  $n_\alpha\cong 8\times 10^{-4}v_0^{-1}$.

We may understand the behavior  of $\sigma_{\rm e}$ 
as a function of $g_1$ and $g_2$ by solving the nonlinear 
Poisson-Boltzmann equation \cite{OnukiJCP}, 
with an interface at $z=0$. 
That is, away from the interface $|z| >\xi$,  
 the normalized potential 
$U(z)\equiv e\Phi(z)/k_BT$ obeys   
\be 
\frac{d^2}{dz^2}U = \kappa_K^2\sinh (U-U_K) 
\en  
in the two phases ($K=\alpha$, $\beta$),  
where  $U_\alpha =U(-\infty)$,  
$U_\beta =U(\infty)$, and 
 $\kappa_K= (4\pi n_Ke^2/\ve_K k_BT)^{1/2}$ 
with $\ve_K= \ve_0+\ve_1 \phi_K$.  
In solving Eq.(3.31) we assume the continuity 
of the electric induction $-\ve d\Phi/dz$ 
at $z=0$ (but this does not hold in the presence 
of interfacial orientation of molecular dipoles, 
as will be remarked  in the summary section). 
 The  Poisson-Boltzmann  
 approximation for $\sigma_{\rm e}$  is of the form, 
\bea
\frac{\sigma_{\rm e}^{\rm \small{PB}}}{k_BT} &=&  
\frac{2n_\alpha}{\kappa_{\alpha}}\bigg 
[ \sqrt{1+b^2+
2b\cosh({\Delta U}/{2})}-b-1\bigg] \nonumber\\
&=&  A_s (n_\alpha/\ell_{{\rm B}\alpha} )^{1/2} .
%(\propto {n_{1\alpha}}^{1/2}).  
\ena 
We should have $\sigma_{\rm e} \cong \sigma_{\rm e}^{\rm \small{PB}} $ 
in the thin interface 
limit $\xi \ll\kappa_K^{-1}$. 
In the first line,  the coefficient 
 $b$ is defined by 
\be 
b= ({\ve_\beta /\ve_\alpha})^{1/2} 
\exp[- ({g_1+g_2})\Delta\phi/4],  
\en 
and   $\Delta U=U_\alpha-U_\beta=
(g_1-g_2)\Delta\phi/2$ is the normalized potential difference 
calculated from Eq.(3.8). 
In the second line, 
 $\ell_{{\rm B}\alpha}= 
e^2/\ve_\alpha k_BT$ is the Bjerrum length in phase $\alpha$. 
The second line  indicates that  
the electrostatic contribution to the surface tension 
 is negative and is of order 
  ${n_\alpha}^{1/2}$ as ${n_\alpha}\to 0$ 
away from the solvent criticality, as 
first predicted by Nicols and Pratt 
for liquid-liquid interfaces \cite{Pratt}. 
Remarkably,  the surface tension 
of air-water interfaces exhibited the 
same behavior at very small 
salt densities (known as the Jones-Ray effect) \cite{Jones}, 
though  it has not yet been  
explained reliably \cite{OnukiJCP}.

In the asymptotic limit 
of antagonistic ion pairs, we assume 
$g_1\ge -g_2 \gg 1$, where   
the coefficient $A_s$ in the second line of eq.(3.32) 
grows   as   
\be 
A_s  \cong  
 \pi^{-1/2} (\ve_\beta/\ve_\alpha)^{1/4}\exp({|g_2|\Delta \phi/4}) 
\en 
 We may also examine  the usual 
case of hydrophilic ion pairs in water-oil, 
where  $g_1$ and $g_2$ are both considerably 
larger than unity. In this case $A_s$ becomes small as  
\bea 
A_s &\cong& 
({\ve_\beta /\pi \ve_\alpha})^{1/2} 
[\cosh(\Delta U/2)-1]\nonumber\\
&&\times \exp[ -(g_1+g_2)\Delta\phi/4].
\ena 
In this case, the electrostatic contribution 
 $-\sigma_{\rm e} (\propto n_\alpha^{1/2})$  in $\sigma$ 
could be detected only at extremely  small salt densities.

Analogously, between 
ionic and nonionic liquids, 
 Aerov {\it el al.} \cite{Aerov} 
calculated the surface tension.   
They showed  that if the affinities of 
cations and anions to  neutral molecules are 
very different, 
the surface tension becomes negative.

{\bf {3.5 Mesophase formation with antagonistic salt}}.
Adding an  antagonistic salt with $\gamma_{\rm p}>1$ 
to water-oil,  we have found 
 instability of one-phase states 
with increasing  ${\chi}$ below Eq.(3.15)  and 
 vanishing of the surface tension 
$\sigma$ with increasing the ion content  
as  in Fig.3.  In such cases, a thermodynamic instability 
is induced   with increasing $\chi$ at a fixed 
ion density $n_0=\av{n_1}=\av{n_2}$,  leading to a mesophase.  
To examine this phase ordering, 
we   performed two-dimensional 
simulations  \cite{Araki,Nara} and 
presented an approximate 
phase diagram  \cite{Nara}. We 
here  present preliminary three-dimensional  results. 
The  patterns  to follow  resemble  those 
 in  block copolymers and 
 surfactant systems \cite{Onukibook,Seul}. 
In our case,  mesophases emerge due to the selective 
solvation and the Coulomb   
interaction without  complex molecular structures. 
Solvation-induced  mesophase formation 
can well be expected in polyelectrolytes 
and mixtures of 
ionic and polar liquids.

%4
\begin{figure}[htbp]
\vspace{-10mm}
\begin{center}
\includegraphics[width=0.66\textwidth, bb= 0 0 646 333]{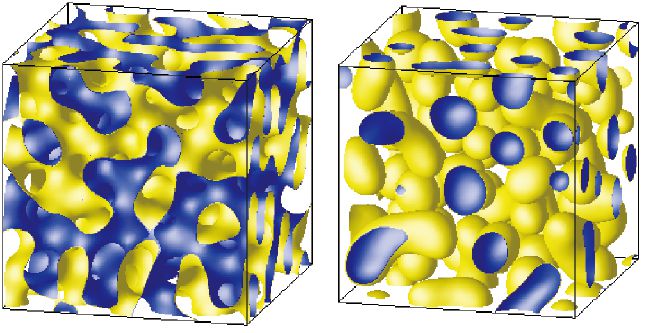}
\caption{Composition  patterns  at $t=5000t_0$ 
for $\chi=2.1$ and $v_0{\bar n} =3\times 10^{-3}$ 
with average composition $\av{\phi}$ being  $0.5$ (left) and 
$0.4$ (right) for antagonistic salt 
with $g_1=-g_2=10$. These patterns are nearly pinned. 
%, $Z_1=-Z_2=1$, $\ell_{\rm B}=3a$ 
Yellow surfaces are oriented to the regions  of $\phi>0.5$ 
and  blue ones are to those of $\phi<0.5$. 
}
\end{center}
\end{figure}

%5
\begin{figure}[htbp]
\vspace{-30mm}
\begin{center}
\includegraphics[width=0.68\textwidth, bb= 0 0 549 398]{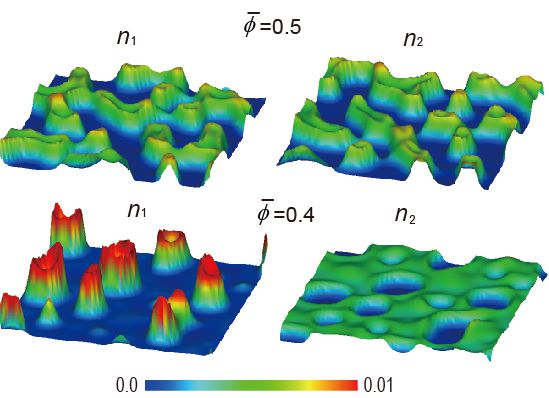}
\caption{
Cross-sectional profiles 
of cation $n_1$ (left) and anion $n_2$ (right) 
in the $x$-$y$ plane at $z=0$ for antagonistic salt 
using data  in Fig. 4. The  domains are 
bicontinuous for  $\av{\phi}=0.5$ (top) 
and droplet-like for $\av{\phi}=0.4$ (bottom).    
}
\end{center}
\end{figure}

%6
\begin{figure}[htbp]
\vspace{-10mm}
\begin{center}
\includegraphics[width=0.5\textwidth, bb= 0 0 631 458]{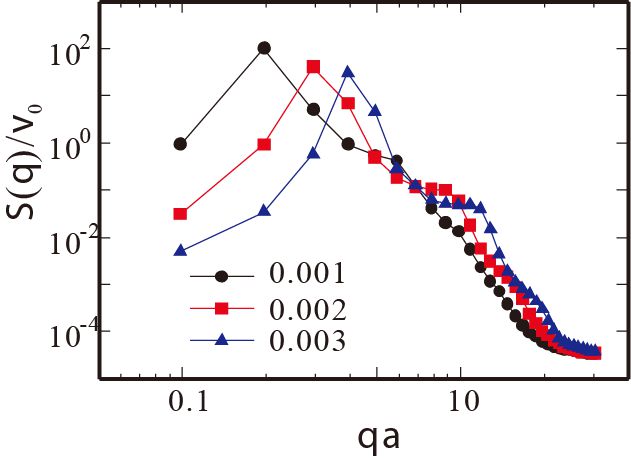}
\caption{Normalized 
structure factor $S(q)/v_0$ for antagonistic salt  
from domain structures in pinned states at $t=5000t_0$, where 
the patterns are bicontinuous at  
for $\av{\phi}=0.5$. A peak height $S(q_m)/v_0$ 
of each curve much exceed unity. 
 The average ion density is  
$v_0n_0=0.001,0.002$, and 0.003 
for the three curves (from above for $q<q_m$). 
}
\end{center}
\end{figure}

We are interested in 
 slow composition evolution with 
antagonistic ion pairs,  
so we assume that the ion distributions are 
given by the modified Poisson-Boltzmann relations in Eq.(3.12). 
The  water composition $\phi$ obeys \cite{Araki,Nara} 
\be
\frac{\partial\phi}{\partial t}+
\nabla\cdot(\phi\mbox{\boldmath $v$})= 
\frac{L_0}{k_BT}  \nabla^2 h,
\en 
where $L_0$ 
is the kinetic coefficient and $h$ is defined by Eq.(3.10). 
Neglecting the acceleration term, we determine 
the velocity field $\bi v$ using  the Stokes approximation,
\be 
\eta_0 \nabla^2{\bi v}
= \nabla p_1+ \phi \nabla h +\sum_i n_i\nabla \mu_i ,
\en 
where $\eta_0$ is the shear viscosity and  
$\mu_i$ are defined by Eq.(3.11), We introduce   $p_1$ to  ensure  
the incompressibility condition 
$\nabla\cdot{\bi v}=0$.  The right hand side of Eq.(3.37) is 
also written as $\nabla\cdot{\tensor \Pi}$, where  
${\tensor \Pi}$ is the stress tensor arising from  
the fluctuations of $\phi$ and $n_i$. 
Here  the total free energy $F$ in Eq.(3.1) 
satisfies $dF/dt \le 0$ with these equations 
(if the boundary effect arising from 
the surface free energy is neglected).

We integrated Eq.(3.36) using the relations 
(3.4), (3.12),  and (3.37) 
on a $64\times 64\times 64$ lattice  
under the periodic boundary condition.  
The system was quenched to an unstable state  
with $\chi=2.1$  at $t=0$.   Space and time 
are measured in units of 
$a=v_0^{1/3}$ and $t_0=a^5/L_0$, respectively. 
Without ions, the diffusion constant 
of the composition 
is given by $L_0 f''/k_BT$ in one-phase states  
in the long wavelength limit (see Eq.(3.14)).
We set  $g_1=-g_2=10$, ${n}_0=3\times 10^{-3}v_0^{-1}$,    
$C=k_BT/2a^2$, $e^2/\ve_0 k_BT =3a$, $\ve_1=0$, 
and $\eta_0L_0/k_BT=0.16a^4$.

In Fig.4, we show  the simulated domain patterns 
at $t=5000t_0$, where we 
can see a bicontinuous structure for $\av{\phi}=0.5$ 
and a droplet structure for $\av{\phi}=0.4$. 
There is almost no further time evolution from this stage.  
In Fig.5, the ion distributions 
are displayed for these two cases 
in the $x$-$y$ plane at $z=0$.  For $\av{\phi}=0.5$, 
the ion distributions are peaked at the interfaces 
forming electric double layers 
(as in the right bottom plate of 
Fig.2).  For  $\av{\phi}=0.4$,  
 the anions are broadly distributed 
 in the percolated oil region, but 
 we expect formation of electric double layers 
with increasing the domain size also in the off-critical condition.  
In Fig.6, the structure factor $S(q)$ in 
steady states are plotted for 
$v_0n_0=0.001,0.002$, and 0.003. 
The peak position decreases with increasing 
$n_0$ in accord with Eq.(3.16). 
Sadakane {\it et al.} \cite{Sadakane,SadakanePRL} 
observed the structure factor similar to those in Fig.6.  

Finally,  we remark that 
the thermal noise, which is absent 
 in our simulation, 
 should be crucial  near the 
criticality of low-molecilar-weight 
solvents. It is  needed 
to explain anomalously  
enhanced composition fluctuations induced by 
NaBPh$_{4}$ near the solvent criticality\cite{Seto}.

\section{Ionic surfactant with  amphiphilic 
and  solvation interactions}
\setcounter{equation}{0}
{\bf {4.1 Ginzburg-Landau theory}}. 
In this section, 
we will give a  diffuse-interface model 
of ionic surfactants\cite{OnukiEPL}, where surfactant 
 molecules are treated as ionized rods. 
 Their  two ends can stay in  very different 
 environments (water and oil) if they are longer than 
 the interface thickness $\xi$.  
In our model,   the 
  adsorption of ionic surfactant molecules    and counterions 
to an oil-water interface 
strongly depends on the selective solvation parameters 
$g_1$ and $g_2$ and that the surface tension 
contains the electrostatic contribution as in Eqs.(3.26) 
and (3.28). 

We add a small amount of cationic surfactant,
anionic counterions in water-oil in the monovalent case.   
The densities of water, oil,   surfactant, and counterion 
are $n_A$, $n_B$, $n_1$, and $n_2$, respectively. 
The  volume fractions of the first three components 
are $\phi_A=v_0n_A$, $\phi_B=v_0 n_B$, and 
$ v_1 n_1$,  where   $v_0$  
is the common molecular volume of 
water and oil  and 
 $v_1$ is the  surfactant molecular volume. 
 The volume ratio  $N_1=v_1/v_0$ can be large, so we 
  do not neglect the surfactant volume fraction, 
  while   we neglect the counterion  volume fraction 
  supposing  a small size  of the counterions.  
 We assume the space-filling condition,
 \be  
\phi_A+\phi_B+ v_1 n_1=1.
\en 
 Let  $2\psi=\phi_A-\phi_B$ be   the 
composition difference between water and oil; then, 
\bea 
&&\phi_A= (1-v_1n_1)/2+\psi, \nonumber\\ 
&&\phi_B= (1-v_1n_1)/2-\psi.
\ena

The   total free energy 
$F$ is again expressed as  in Eq.(3.1).
Similarly to Eq.(3.2), the first part reads  
\bea 
\frac{f_{\rm tot}}{k_BT}  
&=&  \frac{1}{v_0}[\phi_A\ln \phi_A 
 + \phi_B\ln \phi_B  + \chi \phi_A\phi_B] 
\nonumber\\   
&&\hspace{-1cm} +  \sum_i n_i [\ln (n_i\lambda_i^3)- g_i n_i\psi] - 
n_1 \ln Z_{\rm a} ,  
\ena 
The coefficients $g_1$ and $g_2$ are the solvation 
parameters of the ionic surfactant and the counterions, 
respectively. Though a surfactant molecule is amphiphilic, 
it can have preference to water or oil on the average. 
The  last term represents the amphiphilic 
interaction between the surfactant and the 
composition. That is, $Z_{\rm a}$ is the partition function of 
 a rod-like dipole  with its center at the position $\bi r$. 
We assume that the surfactant molecules take a rod-like shape  
with a length $2\ell$  considerably longer than $a=v_0^{1/3}$. 
It is given by the  following integral on  the surface of a sphere 
with radius $\ell$,    
\be 
Z_{\rm a}({\bi r}) = \int\frac{{d\Omega}}{4\pi} 
 \exp \bigg [w_{\rm a} \psi({\bi r}-\ell {\bi u}) - 
w_{\rm a} \psi({\bi r}+\ell {\bi u}) 
\bigg], 
\en 
where  $\bi u$  is the unit vector along the rod direction 
and $\int d\Omega$ represents the integration 
over the angles of $\bi u$. 
The  two ends of the rod  are at 
${\bi r} + \ell {\bi u}$ and 
${\bi r} -\ell {\bi u}$  under the influence of 
 the solvation potentials given by   
$k_BT w_{\rm a} \psi({\bi r}+\ell {\bi u})$  
and 
$-k_B T w_{\rm a} \psi({\bi r}-\ell {\bi u})$. 
The parameter $w_{\rm a}$ represents the strength of the amphiphilic 
interaction.

Adsorption is strong  for large 
$w_{\rm a} \Delta\psi\gg 1$, where  
 $\Delta\psi=\psi_\alpha-\psi_\beta(\cong \phi_{A\alpha}-\phi_{A\beta})$ 
is the difference of $\psi$ between the two phases $\alpha$ 
and $\beta$. In   the one-dimensional case,  
all the quantities vary along the $z$ axis and 
 $Z_{\rm a}$ is rewritten as  
\be 
Z_{\rm a}(z) =    
 \int_{-\ell}^\ell  \frac{ {d\zeta}}{2\ell } \exp [
w_{\rm a} \psi(z-\zeta)-w_{\rm a} \psi(z+\zeta)] .
\en
where  $\zeta=\ell u_z$.  
In the thin interface limit  $\xi\ll\ell$, 
we place  the interface at $z=0$ to find  $Z_{\rm a} = 1$  
for $|z|>\ell$, while       
\be 
 Z_{\rm a} (z) \cong 1+ ({1-|z|/\ell}) [\cosh({w_{\rm a} \Delta\psi})-1],
\en 
for $|z|<\ell$.  Furthermore,  
in the dilute limit $v_1n_1\ll 1$ and without the electrostatic 
interaction, 
we have $n_1(z)= n_{1\alpha}Z_{\rm a}(z)$ for $z<0$ 
and  $n_1(z)= n_{1\beta}Z_{\rm a}(z)$ for $z>0$, where 
$n_{1\alpha}$ and $n_{1\beta}=e^{-g_1\Delta\psi}n_{1\alpha}$  
are the bulk surfactant densities. 
The surfactant adsorption  then grows as 
\bea 
\Gamma_1 &&=\int_{-\infty}^0 dz[n_1(z)-n_\alpha]
+  \int_0^{\infty} dz[n_1(z)-n_\beta]\nonumber\\  
&&= (n_{1\alpha}+n_{1\beta})\ell  
[\cosh({w_{\rm a} \Delta\psi})-1]/2.
\ena  
%which   grows for ${w_{\rm a} \Delta\psi}\gg 1$. 
However,  the steric effect 
comes into play at the interface 
with increasing 
the surfactant volume fraction at the interface ($\sim \Gamma_1 v_1/\ell$).

%7
\begin{figure}[htbp]
\vspace{-40mm}
\begin{center}
\includegraphics[scale=0.83,bb=0 0 331 530]{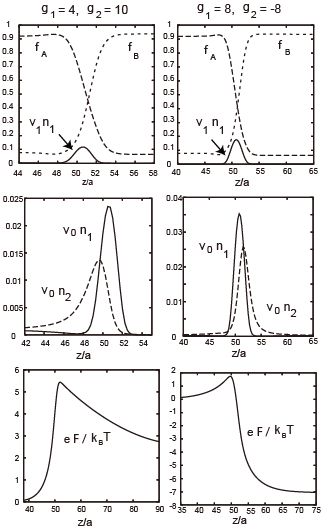}
\caption{ Profiles for mixtures with 
cationic surfactant and anionic counterions 
with  $v_1=5v_0$ and $w_{\rm a}=12$. 
Top: $v_1n_1$ (bold line), $\phi_A$, and $\phi_B$. Middle:  
$v_0n_1$ and $v_0n_2$. Bottom: 
$e\Phi/k_BT$ exhibiting a maximum at the interface.  Here  
$g_1=4$, $g_2=10$, and   $v_0n_{1\alpha}=10^{-3}$ (left), 
while  $g_1=-g_2=8$ and  $v_0n_{1\alpha}=
3.6\times 10^{-4}$ (right). 
The counterion distribution 
has a peak in the phase $\alpha$ (left) 
or $\beta$ (right) depending on $g_2$. 
$[$From:  A. Onuki, Europhys. Lett. 
 {\bf 82}, 58002 (2008)$]$.
}
\end{center}
\end{figure}

%8
\begin{figure}[htbp]
\vspace{-30mm}
\begin{center}
\includegraphics[scale=0.9, bb= 0 0 237 384]{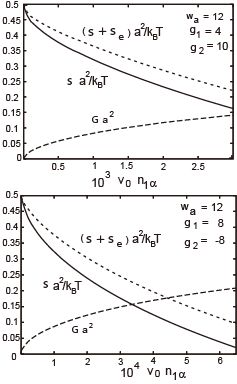}
\caption{
$\sigma a^2/k_BT$, $(\sigma+\sigma_{\rm e}) a^2/k_BT$,  and   
$\Gamma a^2$ as functions of $v_0n_{1\alpha}$ 
with  $v_1=5v_0$, $\chi=3$,  and $w_{\rm a}=12$.  
The curves change on a scale of 
$10^{-3}$ for hydrophilic ion pair 
$g_1=4$ and $g_2=10$ (top) 
and on a scale of 
$10^{-4}$ for antagonistic ion pair 
$g_1=-g_2=8$ (bottom).
$[$From:  A. Onuki, Europhys. Lett. 
 {\bf 82}, 58002 (2008)$]$.}
\end{center}
\end{figure}

{\bf {4.2  Interface profiles of compositions, 
ion densities, and potential}}. 
We give typical  one-dimensional interface profiles 
 varying along the $z$-axis in Fig.7. 
We set  $v_1=5v_0$, $C= 3 k_BT /a$, $\chi=3$, 
 and  $e^2 /a\ve_ck_BT= 16/\pi$. 
The dielectric constant is assumed to be 
of the form   $\ve= \ve_c(1+ 0.8\psi)$, where 
$\ve_c$ is the critical value.
Then  $\ve_\alpha \cong 
2\ve_\beta$  at $\chi=3$. This figure 
was produced in the presence of 
 the image interaction   in our previous work\cite{OnukiEPL}
(though it is not essential here).

In Fig.7,  we show the volume fractions  $\phi_A$, $\phi_B$, 
and $v_1 n_1=1-\phi_A-\phi_B$ (top),   
the ion densities $n_1$ and $n_2$ (middle), 
and the potential $e\Phi/k_BT$ with $\Phi_\alpha=0$ (bottom).   
In the  left, the counterions are 
more hydrophilic than the cationic surfactant, where 
 $g_1=4$ and  $g_2=10$ leading to 
  $\Gamma=0.124 a^{-2}$ and $\sigma=0.317 k_BTa^{-2}$ at 
$n_{1\alpha}= 10^{-3}v_0^{-1}$. 
In the  right  plates, the  surfactant cations are 
hydrophilic and the counterions 
are hydrophobic, where      $g_1=-g_2=8$ leading to 
$\Gamma= 0.155 a^{-2}$ 
and $\sigma  = 0.159 k_BTa^{-2}$ 
 at $n_{1\alpha}=3.6\times 10^{-4}v_0^{-1}$.  
  The distribution of the   surfactant $n_1$ 
is narrower than that of the counterions $n_2$. 
This gives rise  to a peak of   $\Phi$ at $z=z_{\rm p}$, 
at which $E(z_{\rm p})\propto 
\int_{-\infty}^{z_{\rm p}}dz (n_1(z)-n_2(z))=0$.

The adsorption  strongly depends on 
the solvation parameters $g_1$ and $g_2$.
It is much more enhanced 
for antagonistic ion pairs  than for hydrophilic 
 ion pairs.

{\bf {4.3  Surface tension}}. 
The  grand potential density 
$\omega$ is again given by Eq.(3.24) 
and tends to a common constant 
$\omega_\infty$ as $z \to \pm \infty$, 
though its form is more complicated. 
The surface tension 
$\sigma=\int dz[\omega(z)-\omega_\infty]$ 
is rewritten as Eq.(3.26) 
and is approximated as Eq.(3.28) for small $n_{1\alpha}$. 
The areal electrostatic-energy density  $\sigma_{\rm e}$ 
 in Eq.(3.27) is again important in the present case.

In Fig.8, we show $\sigma$, $\sigma+\sigma_{\rm e}$, 
and $\Gamma$ 
as functions of $v_0n_{1\alpha}$ at $w_{\rm a}=12$, 
where $\Gamma$ is defined as in Eq.(3.28) 
for $n=n_1+n_2$.   In the upper plate, 
the two species of ions  
are both hydrophilic ($g_1=4$ and $g_2=10$), 
while in the lower plate 
the surfactant and the counterions 
are antagonistic ($g_1=8$ and $g_2=-8$). 
For the latter case, 
a large electric double layer is formed 
at an  interface,  leading to  
a   large $\Gamma$ and a large decrease in $\sigma$ 
even at very small $n_{1\alpha}$. 
In the present  case   
the Gibbs term $k_BT \Gamma$ 
is a few times larger than $\sigma_{\rm e}$. 
Note that the approximate formula 
(3.28) may be derived also in this case, but 
it  is valid only for very small $n_{1\alpha}$ 
in Fig.8.

\setcounter{equation}{0}
\section{Phase separation due to strong 
selective solvation}
{\bf {5.1 Strongly hydrophilic or hydrophobic solute}}. 
With addition of a strongly selective solute in  
a binary mixture in one-phase states, 
we predict  precipitation of domains 
composed of  the preferred  component enriched 
with the solute\cite{Okamoto}.  
These precipitation phenomena 
occur both for a hydrophilic salt (such as NaCl)  and 
a neutral hydrophobic solute \cite{S1,S2,S3,S4,S5,S6,S7}. 
In our scheme, a   very large size of the selective 
solvation parameter $g_i$ is essential. 
In Secs.2 and 3,  we 
have shown  that $|g_i|$ can well exceed 
$10$ both for hydrophilic and hydrophobic solutes.

With  hydrophilic cations and anions,   a  charge 
density  appears  only near the 
interfaces, shifting  the surface tension  slightly. 
Thus, in the static aspect of precipitation, 
the electrostatic interaction is not essential, 
while  fusion of precipitated domains should be 
suppressed by the presence of the electric double layers. 
We  will first treat  a hydrophilic neutral solute as a third 
 component, but the following results 
are applicable also to  a neutral   hydrophobic solute  
  if  water and oil are exchanged.  
In addition, in a numerical example in Fig.11, we will include 
the electrostatic interaction among hydrophilic ions.

{\bf {5.2 Conditions of two phase coexistence}}. 
Adding  a small amount of a highly 
selective solute in  water-oil, we assume the 
 following free energy density,  
\be 
{f}_{\rm tot} (\phi,n)  =  f(\phi) +k_BT{n}[\ln (n\lambda^3) 
 -1- g \phi].    
\en 
This is a general model  for a dilute solute. 
For monovalent electrolytes,  this form 
follows from Eq.(3.11)  if there is  no charge density 
or if we set 
\be 
n_1=n_2=n/2, \quad g=(g_1+g_2)/2,
\en 
The first term $f(\phi)$  is assumed to be 
of the Bragg-Williams  form (3.3).   
The $\lambda$ is the thermal de Broglie length.  
The solvation  term ($\propto g$) 
 arises from the solute  preference  of  water over oil 
 (or oil over water). 
The strength $g$  is 
assumed to much exceed unity\cite{OnukiPRE}.   
We  fix  the  amounts   
of the constituent components in the cell with a volume $V$. 
Then the   averages   
$ \bar\phi =\av{\phi}= \int d{\bi r} \phi/V$ and 
${\bar n}=\av{n}= \int d{\bi r} n/V$ are given 
 control parameters as well as  $\chi$.

In   two phase coexistence in equilibrium, let 
the composition and the solute density be 
$(\phi_\alpha,n_\alpha)$ in  phase $\alpha$ 
and  $(\phi_\beta,n_\beta)$ 
in  phase $\beta$, where  $\phi_\alpha>{\bar \phi}
>\phi_\beta$ and $n_\alpha>{\bar n}
>n_\beta$.  We introduce  the  chemical potentials 
 $h= {\p f_{\rm tot}}/{\p \phi}$ and 
 $\mu= {\p f_{\rm tot}}/{\p n}$. Equation (5.1) yields  
\bea 
&&h=f'(\phi) -k_BTgn, \\ 
&&\mu/k_BT=\ln (n\lambda^3)- g\phi,
\ena  
where $f'=\p f/\phi$. 
The system is 
linearly stable for  
$ 
\p h/\p \phi- (\p h/\p n)^2/\p \mu/\p n>0$ 
 or for 
\be 
f''(\phi) - k_BTg^2 n>0,
\en  
where $f''= \p^2 f/\p \phi^2$. 
Spinodal decomposition occurs 
if the left hand side of Eq.(5.5) is negative.

The  homogeneity 
of $\mu$  yields  
\be 
n= {\bar n} e^{g\phi}/\av{e^{g\phi}},   
\en 
where $\av{\cdots}$ denotes taking the space average. 
The bulk solute densities 
are $n_K= {\bar n} e^{g\phi_K}/\av{e^{g\phi}}$  
for  $K=\alpha, \beta$ in two-phase coexistence. 
In our approximation Eq.(5.6) holds even in the interface 
regions.  We write the  volume fraction of 
 phase $\alpha$ as  $\gamma_\alpha$. We then have  
$
\av{e^{g\phi}}= \gamma_\alpha e^{g\phi_\alpha} 
 +(1-\gamma_\alpha) e^{g\phi_\beta}$ in Eq.(5.6).
In terms of 
$\bar\phi$ and $\bar n$, $\gamma_\alpha$  is expressed   as 
\be
\gamma_\alpha 
= ({{\bar \phi}-\phi_\beta})/{\Delta \phi}
=  ({{\bar n}-n_\beta})/{\Delta n}.  
\en  
where $\Delta \phi=\phi_\alpha-\phi_\beta>0$ 
and $\Delta n=n_\alpha-n_\beta>0$. 
In these expressions we neglect the volume of the 
 interface regions.  
Since $n_\alpha/n_\beta = e^{g\Delta\phi}\gg 1$ 
for  $g\Delta \phi\gg 1$,      
the solute  is much more enriched  in 
 phase $\alpha$ than in phase $\beta$.  
Eliminating   $n$  using   Eq.(5.6), 
 we may express the 
average free energy density  as 
\be 
\av{f_{\rm tot}}
=
 \av{f} - k_B T{\bar n} \ln [\av{e^{g\phi}}]
+A_1,    
\en 
where $A_1= k_B T{\bar n}
 [\ln ({\bar n}\lambda^3)-1]$ is a constant at fixed $\bar n$. 
In terms of $\phi_\alpha$, $\phi_\beta$, and $\gamma_\alpha$, 
Eq.(5.8)  is rewritten as 
\bea 
\av{f_{\rm tot}}
&=& [\gamma_\alpha f(\phi_\alpha) 
+ (1-\gamma_\alpha) f(\phi_\beta)]   \nonumber \\
&&\hspace{-1cm} 
- k_B T{\bar n} \ln [ 
 \gamma_\alpha e^{g\phi_\alpha} 
 +(1-\gamma_\alpha) e^{g\phi_\beta}] +A_1. 
\ena 
The second   term ($\propto {\bar n})$  is  
 relevant for large $g$ (even for small $\bar n$). 
Now we  should minimize 
$\av{f_{\rm tot}}-h[\gamma_\alpha\phi_\alpha 
+ (1-\gamma_\alpha)\phi_\beta -\bar{\phi}]$ 
with respect to $\phi_\alpha$, 
$\phi_\beta$, and $\gamma_\alpha$ at fixed $\bar\phi$,  
where $h$ appears as the Lagrange multiplier. 
 Then 
we obtain the equilibrium 
conditions of two-phase coexistence,   
\bea
&&\hspace{-3mm} h=f'(\phi_\alpha) -k_BTgn_\alpha 
 = f'(\phi_\beta) -k_BTgn_\beta, \\
&& {f(\phi_\alpha)-f(\phi_\beta)}- k_BT\Delta n=h{\Delta\phi}. 
\ena   
These static relations hold even for ion pairs 
under Eq.(5.2).
%  if a  charge density 
%appears only near the interfaces. 

%9
\begin{figure}[htbp]
\vspace{-30mm}
\begin{center}
\includegraphics[scale=0.48, bb= 0 0 657 686]{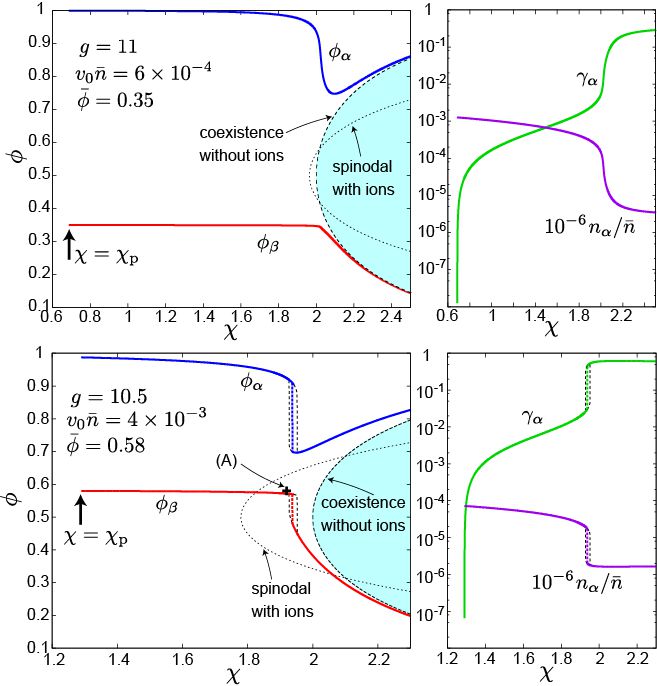}
\caption{
Left: compositions $\phi_\alpha$ and $\phi_\beta$ vs $\chi$. 
Right: semi-logarithmic plots of 
volume fraction  $\gamma_\alpha$  
and normalized solute density 
$n_\alpha/{\bar n}$   of the water-rich phase $\alpha$ vs $\chi$.   
Here  $g=11$, ${\bar \phi}=0.35$, and 
 ${\bar n}= 6\times 10^{-4}v_0^{-1}$ (top), 
 while  $g=10.5$, ${\bar \phi}=0.58$, 
  and ${\bar n}= 4\times 10^{-3}v_0^{-1}$ (bottom). 
 Plotted in the left  
 also are the coexisting  region without solute (in right green) 
and the spinodal curve with solute (on which  the 
left hand side of  Eq.(5.5) vanishes). See Fig.11 for simulation of 
phase separation on  point (A) 
in the left bottom panel.
$[$Upper plates are from:  
R. Okamoto and A. Onuki,  Phys. Rev. E {\bf 82},
 051501 (2010).$]$} 
\end{center}
\end{figure}

{\bf {5.3 Numerical results  of two phase coexistence}}. 
In Fig.9,  we give  numerical results 
 on  the phase behavior  
 of   $\phi_\alpha$ and  
 $\phi_\beta$ in the left   
 and  $n_\alpha$ and 
  $\gamma_\alpha$ in the right   
  as functions of $\chi$. 
  We set  $g=11$,    $\bphi=0.35$ and ${\bar n}= 6\times 
  10^{-4}v_0^{-1}$ in the top plates 
 and   $g=10.5$, ${\bar \phi}=0.58$, 
  and ${\bar n}= 4\times 10^{-3}v_0^{-1}$ 
  in the bottom plates. The solute density is much larger 
  in the latter case. Remarkably, a precipitation branch appears in the range, 
\be  
\chi_{\rm p}  < \chi< 2.
\en
The  volume fraction 
$\gamma_\alpha$ decreases to zero 
as $\chi$ approaches  the  lower bound  
$\chi_{\rm p}=\chi_{\rm p} ({\bar \phi},{\bar n})$. 
Without solute, the 
 mixture  would be in one-phase states for $\chi<2$. 
The precipitated domains are  
solute-rich  with 
$\phi_\alpha \cong 1$, while $\phi_\beta$ is slightly 
larger than $\bar\phi$. In the left upper plate 
$\phi_\alpha$ increases continuously with 
decreasing $\chi$, while in the left lower plate 
$\phi_\alpha$ jumps at $\chi=1.937$ 
and hysteresis appears in the region 
$1.927< \chi < 1.952$.
We also plot the spinodal 
curve,
$
f''({\bar\phi}) - k_BTg^2 {\bar n}=0,
$  
following from Eq.(5.5). 
 Outside this curve,  
homogeneous states are metastable and precipitation 
can proceed  via 
homogeneous nucleation in the bulk or 
via heterogeneous nucleation 
on hydrophilic surfaces  of boundary plates or 
colloids \cite{Okamoto}. Inside this curve, 
the system is linearly unstable and 
precipitation occurs via spinodal decomposition.
This unstable region is expanded 
for  ${\bar n}= 4\times   10^{-3}v_0^{-1}$ 
in the lower plate.  
  
%10
\begin{figure}[htbp]
\vspace{-15mm}
\begin{center}
\includegraphics[scale=0.46, bb=0 0 712 344]{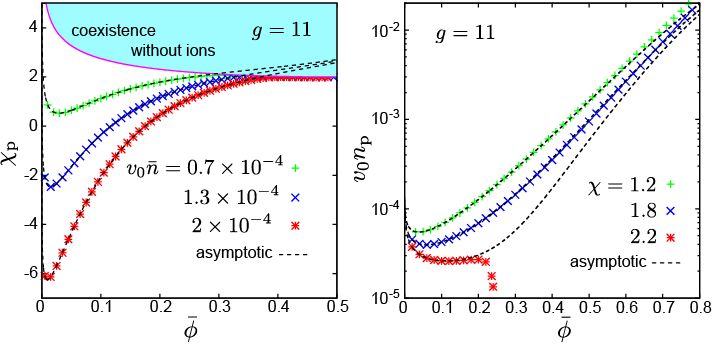}
\caption{
 Left:$\chi_{\rm p}(\bar{\phi},{\bar n})$ 
vs $\bar \phi$ for  three values of 
${\bar n}$ 
%$5\times 10^{-6}$, and  $10^{-5}$ from above  
at $g=11$,  which nearly coincide 
with the asymptotic formula (5.22) 
(dotted line) for $\bar{\phi} <0.35$ and 
converge to the coexistence curve for larger $\bar\phi$. 
Coexistence region  without  ions is in the upper region 
(in right green).  Right: $v_0 n_{\rm p}(\bar{\phi},\chi)$ 
vs $\bar \phi$  on  a semi-logarithmic scale 
%for  $\chi=1$, 1.5, and 2.3 from above 
 at $g=11$, 
It coincides with the asymptotic formula (5.23) 
(dotted line) for $\chi<2$ and  tends  to zero at  
the coexistence composition,  0.249   (top)  
or  0.204  (bottom). 
$[$From: R. Okamoto and A. Onuki,  Phys. Rev. E {\bf 82},
 051501 (2010).$]$}
\end{center}
\end{figure}

{\bf {5.4 Theory of asymptotic behavior for large $g$}}. 
We present  a theory  of  
the precipitation branch  in  the limit 
 $g\gg 1$ to determine $\chi_{\rm p}$ and $n_{\rm p}$. 
 Assuming  the branch (5.12) 
 at the starting point, we 
 confirm its existence  self-consistently. 
% Since $\phi_\alpha\cong 1$, 
% the logarithmic term $(\propto (1-\phi)\ln(1-\phi)$) 
% in the free energy density (3.3)  is crucial 
% in phase $\alpha$. 

We first  neglect the term   $-k_BTgn_\beta$ in Eq.(5.10) 
from $gv_0 n_\beta \ll 1$ and the term 
$f(\phi_\alpha)$ in Eq.(5.11) from 
$\phi_\alpha \cong 1$. 
In fact  $gv_0 n_\beta\ll 1$ in Fig.9. 
We then obtain   
\be 
h \cong   
f'(\phi_\beta)\cong 
- [{f(\phi_\beta)+k_BTn_\alpha}]/({1-\phi_\beta}) . 
\en 
This determines the  solute density $n_\alpha$ 
in  phase $\alpha$ 
as a function of $\phi_\beta$ in the form,  
\be 
n_\alpha\cong G(\phi_\beta)/k_BT,
\en 
where $G(\phi)$ is a function of $\phi$  defined as  
\bea 
G(\phi) &=& -f(\phi) - (1-\phi)f'(\phi)\nonumber\\
&=& - (k_BT/v_0) [\ln \phi+ \chi(1-\phi)^2].
\ena 
From $dG/d\phi=  - (1-\phi)f''(\phi)<0$ 
and $G(1)=0$, we have 
$G(\phi)>0$  outside the 
coexistence curve, ensuring $n_\alpha>0$ in Eq.(5.14).

In Eq.(5.10),   we  next 
use Eq.(3.3) for $f'(\phi_\alpha)$ to obtain 
\be 
 v_0^{-1}
 k_BT [-\ln (1-\phi_\alpha)-\chi- 
 g v_0n_\alpha] \cong f'(\phi_\beta),
\en 
  where  the logarithmic term ($\propto \ln(1-\phi)$) 
balances with the solvation term   ($\propto gn_\alpha$).  
Use of  Eq.(5.15) gives    
\be 
 {1-\phi_\alpha}\cong A_\beta \exp[-g v_0G(\phi_\beta)/k_BT],
\en 
where the coefficient $A_\beta$  is given by 
\be 
A_\beta=\exp[-\chi - v_0 f'(\phi_\beta)/k_BT],
\en 
so $A_\beta$ is of order unity.  
The factor $\exp[-g v_0 G(\phi_\beta)/k_BT]$  
in Eq.(5.17) is very small for $g\gg 1$, 
leading to $\phi_\alpha \cong 1$.

Furthermore, from Eqs.(5.6) and (5.7), 
the volume fraction $\gamma_\alpha$ 
of phase $\alpha$ is approximated as  
\be 
\gamma_\alpha 
%({\bar \phi}-\phi_\beta)/\Delta\phi
\cong 
{\bar n}/n_\alpha -e^{-g\Delta \phi}\cong 
k_BT{\bar n}/G(\phi_\beta) -e^{-g\Delta \phi}.
%= v_0{\bar n}/G(\phi_\beta)-e^{-g\Delta \phi}$.   
\en   
The above relation is rewritten as 
\bea  
&&\hspace{-1.2cm}G(\phi_\beta) \cong  {k_BT{\bar n}}/{
(\gamma_\alpha+  e^{-g\Delta\phi})}\nonumber\\
&&\hspace{-2mm}\cong\frac{k_BT{\bar n}(1-\phi_\beta)}{
\bar{\phi}-\phi_\beta+  (1-\phi_\beta)
\exp[{-g(1-\phi_\beta)]}}. 
\ena 
From the first to  second line,  we have used   Eq.(5.7)  and 
 replaced  $\Delta\phi$ by $ 1-\phi_\beta$. This  equation 
  determines  $\phi_\beta$ and   $\gamma_\alpha
\cong ({\bar \phi}-\phi_\beta)/(1-\phi_\beta)$.  
 We  recognize that  
 $G(\phi_\beta)$ increases  up to $T{\bar n}e^{g\Delta\phi}
\cong  T{\bar n}e^{g(1-{\bar\phi})}$  as  
$\gamma_\alpha \to 0$ or as $\phi_\beta \to \bar\phi$. 
In this  limit  it follows   the marginal relation, 
\be 
G({\bar \phi})\cong k_B T{\bar n}
e^{g(1-{\bar\phi})} \quad (\gamma_\alpha\to 0).
\en 
If ${\bar n}$ is fixed, 
this relation   holds 
at ${\chi}=\chi_{\rm p}$ so that 
\be 
\chi_{\rm p} \cong 
 [-\ln \bphi- v_0{\bar n}e^{g(1-\bphi)}]/(1-\bphi)^2,  
\en 
where we use the second line of  Eq.(5.15). 
Here $\bar n$ appears in the combination 
${\bar n}e^{g(1-\bphi)}(\gg {\bar n})$.  
On the other hand, if $\chi$ is fixed, Eq.(5.21)  holds 
at ${\bar n}=n_{\rm p}$. Thus the minimum solute density 
$n_{\rm p}$ is estimated as 
\be 
n_{\rm p} \cong  e^{-g(1-{\bar \phi})}G({\bar \phi})/k_BT, 
\en
which  is much decreased 
by the small factor $e^{-g(1-{\bar \phi})}$.

In Fig.10, the  curves of $\chi_{\rm p}$ and $n_{\rm p}$ 
 nearly coincide with  the 
 asymptotic formulas (5.22) and (5.23) in the range 
 ${\bar\phi}<0.35$ 
for $\chi_{\rm p}$  and  in the range 
 $\chi<2$ for ${n}_{\rm p}$. They  exhibit a minimum 
 at $\bphi \sim 
e^{-g}/v_0{\bar n}g$ for $\chi_{\rm p}$ 
and  at $\bphi \sim 
g^{-1}$ for ${n}_{\rm p}$. For larger $\bar{\phi}>0.35$,  
$\chi_{\rm p}$ nearly  coincide with the coexistence curve, 
indicating  disappearance of the precipitation branch. 
 Notice that   ${n}_{\rm p}$ 
decreases to   zero as $\bar\phi$ approaches 
the coexistence composition  $\phi_{\rm cx}
= 0.249$   at  $\chi=2.2$ (top)  
and 0.204  at  $\chi=2.3$ (bottom), 
where phase separation occurs without solute.

{\bf {5.5 Simulation of spinodal decomposition for hydrophilic 
ions}}. In our theory \cite{Okamoto} 
we investigated solute-induced 
nucleation starting with  homogeneous metastable states 
outside the spinodal curve (dotted line) in the left panels 
of Fig.9. Here, we show two-dimensional numerical results of 
spinodal decomposition for hydrophilic ions with $g_1=12$, 
 $g_2=9$, $e^2/\ve_0k_BT=3a$, and $\ve_1=\ve_0$.  At $t=0$, 
we started  with point (A) inside the spinodal curve 
in the left lower panel in  Fig.9 using  
the common values of the static 
parameters  given by $\chi=1.92$, 
 ${\bar \phi}=0.58$,  and ${\bar n}=\av{n_1+n_2}=
   4\times 10^{-3}/v_0$.  In addition, we set $C=2k_BTa^2/v_0$. 
On a $256\times 256$ lattice under the periodic 
boundary condition, we integrated Eq.(3.36) for the composition 
$\phi({\bi r},t)$  with the velocity field ${\bi v}({\bi r},t)$ 
being determined by the Stokes approximation 
in Eq.(3.37). The cations and anions obey 
\bea 
&&\frac{\partial n_i}{\partial t}+
\nabla\cdot(n_i\mbox{\boldmath $v$})
= \frac{D}{k_BT}\nabla \cdot n \nabla \mu_i \nonumber\\ 
&&=D\nabla\cdot\bigg[{\nabla{n_i}} -
g_in_i\nabla\phi -Z_ie n_i{\bi E}\bigg],
\ena 
where $i=1,2$. The chemical 
potentials $\mu_i$ are  defined in Eq.(3.11) 
and the  ion diffusion constants are commonly  given by $D$. 
The space mesh size is $a= v_0^{1/d}$ with 
$d=2$.   We measure  time  in units of  
$t_0 =a^2v_0 / L_0$ and  set $D=a^2/ t_0$ 
and $\eta_0 =0.15 k_BT t_0 / v_0$, 
where $L_0$ is the kinetic coefficient  
 for the composition   and   $\eta_0$ is   
 the shear viscosity.

In Fig.11, we show the time evolution of the 
droplet volume fraction, which is the fraction 
of the region $\phi>0.6$.   
In the early stage, the droplet number decreases 
in time  with 
the evaporation and condensation mechanism. 
In the late stage, it changes  
very slowly tending to a constant. 
In our simulation  without  random noise, 
the droplets do not undergo Brownian motion and 
the droplet collision is suppressed. 
We also performed a  simulation for a neutral 
solute with $g=10.5$ 
(not shown here), which exhibits  almost 
the same phase separation 
behavior as in  Fig.11.

%11
\begin{figure}[htbp]
\vspace{-20mm}
\begin{center}
\includegraphics[scale=0.65, bb= 0 0 490 367]{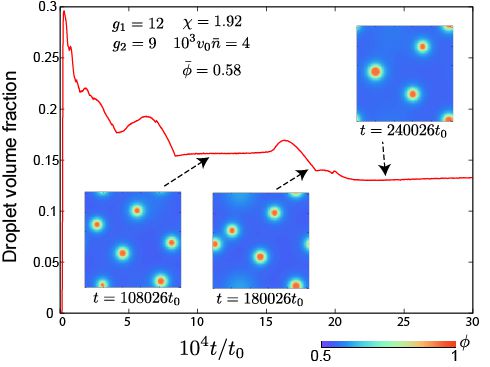}
\caption{Time evolution of  droplet 
volume fraction (that of the region 
$\phi>0.6$)  and composition snapshots of  
precipitated droplets at three times. 
They are  induced by hydrophilic ions  
with  $g_1=12$ and $g_1=9$. 
The initial state was at point (A) in Fig.9  
with $\chi=1.92$,  ${\bar \phi}=0.58$, 
  and ${\bar n}= 4\times 10^{-3}v_0^{-1}$.  
The droplet volume fraction nearly tends to  a constant.      
}
\end{center}
\end{figure}

\setcounter{equation}{0}
\section{Theory of  polyelectrolytes} 
{\bf {6.1 Weakly ionized polyelectrolytes}}. 
In this section, 
we consider weakly charged polymers  in a theta or poor,  
one-component water-like  solvent 
in the semidilute case $\phi>N^{-1/2}$. 
Following the literature of polymer physics \cite{PG}, 
we use  $\phi$ and $N$ to represent 
the polymer volume fraction and 
 the polymerization index.  Here charged particles  
 interact differently  between 
 uncharged monomers and  solvent molecules. 
 The selective solvation should become 
 more complicated for mixture solvents, as discuused 
 in Sec.1.

To ensure flexibility of the chains, 
we assume that the fraction of 
charged monomers on the chains, denoted by  
$f_{\rm ion}$,  is small or $f_{\rm ion}\ll 1$.   
From the scaling theory\cite{PG}, 
the polymers consist of blobs with 
monomer number $g_b=\phi^{-2}$ with  
length $\xi_b = ag_b^{1/2}=a\phi^{-1}$. 
The electrostatic energy within a blob 
is estimated as 
\be 
\epsilon_b= k_BT(f_{\rm ion}g)^2\ell_B/\xi_b= 
k_BTf_{\rm ion}^2\ell_B/\phi^{3}a.
\en  
where $\ell_B$ is the Bjerrum length. The blobs 
are not much deformed under 
the  weak charge condition $\epsilon_b< k_BT$, 
which is rewritten as  
\be 
\phi> f_{\rm ion}^{2/3}(\ell_B/a)^{1/3}.
\en 

{\bf {6.2 Ginzburg-Landau theory}}. 
The number of the ionizable monomers 
(with charge $-e$) on a chain 
is written as $\nu_M N$ with $\nu_M < 1$. 
Then the  degree of ionization (or dissociation)  
 is 
$\zeta= 
f_{\rm ion}/\nu_M  
$ 
and the number density of the ionized monomers is 
\be 
n_p= v_0^{-1}f_{\rm ion}  \phi= 
v_0^{-1}\nu_M  \zeta \phi,
\en 
The charge density is  expressed as 
\be 
\rho=e\sum_{i=c,1,2}Z_in_i -e n_p.
\en  
Here $i=c$ represents  the counterions,  
$i=1$ the added cations,  
and $i=2$ the added anions.  
The required relation  $f_{\rm ion}\ll 1$
 becomes $\nu_M\zeta\ll 1$ (which is 
 satisfied for any $\zeta$ if $\nu_M\ll 1$).

We set up the free energy $F$ 
accounting  for the 
molecular interactions and the ionization equilibrium\cite{Onuki-Okamoto}.
%, neglecting  the image 
%interaction \cite{Onsager,Levin-Flores}  and the  formation  
%of   
Then  $F$ assumes  the standard form (3.1), 
where  the coefficient of the gradient free energy is written as
\cite{Onukibook,PG} 
\be 
C(\phi)=k_BT /12a \phi(1-\phi),
\en 
in terms of the molecular length  $a=v_0^{1/3}$ 
and $\phi$.
The $f_{\rm tot}$ consists of four parts as 
\bea 
{f_{\rm tot}}&=& {f(\phi)}+ {k_BT} 
\sum_{i=c,1,2} {n_i} [\ln (n_i\lambda_i^3)-1 +g_i\phi] 
  \nonumber\\
&&\hspace{-0.1cm}
+{k_BT}(\Delta_0+g_p\phi )n_p + f_{\rm dis}. 
\ena   
The first term $f$ is of 
the Flory-Huggins form \cite{PG,Onukibook},  
\be 
\frac{v_0f}{k_BT}  =  
 \frac{\phi}{N}  \ln\phi + (1-\phi)\ln (1-\phi) 
+ \chi \phi (1-\phi) . 
\en 
The coupling terms ($\propto g_i, g_p$)  
arise from the molecular interactions 
 among the charged  particles (the ions and the charged monomers) and  
 the uncharged particles (the solvent particles and 
 the uncharged monomers), while   
$k_BT\Delta_0$ 
is the dissociation  free energy  in the dilute limit 
of polymers ($\phi \to 0$). 
The last term in $f_{\rm tot}$ arises from  the dissociation 
entropy on chains \cite{Joanny,Bu1,Bu2},  
\be 
\frac{v_0}{k_BT} 
f_{\rm dis}= 
{\nu_M}\phi\bigg 
[\zeta\ln\zeta+(1-\zeta)\ln(1-\zeta)\bigg].
\en

{\bf {6.3 Dissociation  equilibrium}}. 
If $F$ is minimized with respect to $\zeta$, 
it follows   the equation of ionization equilibrium 
or the mass action law,  
\be 
n_c{\zeta}/({1-\zeta}) =K(\phi),   
\en 
where $n_c$ is the counter ion density and $K(\phi)$ is the 
 dissociation constant of the form, 
\be 
K(\phi) =v_0^{-1} \exp[-\Delta_0-(g_p+g_c)\phi]. 
\en 
We may interpret   $k_BT[\Delta_0+ (g_p+g_c)\phi]$ 
as  the composition-dependent  
dissociation free energy.  
  With increasing the polymer volume fraction $\phi$, 
  the dissociation decreases 
for positive $g_p+g_c$ and increases for negative $g_p+g_c$.
If $g_p+g_c \gg 1$, $K(\phi)$ much decreases 
even for a small increase of $\phi$. Here 
$K(\phi)$ has the meaning of the crossover counterion density 
since $\zeta$ is expressed as 
\be 
\zeta= {1}/[1+n_c/K(\phi)],
\en 
which decreases appreciably  for $n_c > K(\phi)$.

%{\it Relations in bulk without salt}. 
In particular, if there is no charge density and no salt 
($n_p=n_c$ and $n_1=n_2=0$),  $n_c$ satisfies 
 the quadratic equation   
$
n_c({n_c+K})=v_0^{-1}\nu_M\phi K,
$
which is solved to give  
\be 
\zeta= {v_0n_c}/{\nu_M\phi} 
= {2}/({\sqrt{Q(\phi) +1}+1}) .
\en 
Here it is convenient to introduce 
\be 
Q(\phi) ={4\nu_M\phi}/{v_0 K(\phi)}.
\en 
We find 
  $\zeta\ll 1$ and $n_c\cong (\nu_M\phi K/v_0)^{1/2}$ 
for $Q \gg 1$, while 
$\zeta \rightarrow 1$   for $Q \ll 1$.
The relation (6.12) holds approximately  
for small charge densities without salt.

{\bf {6.4 Structure factor}}. 
As in Sec.3, it is straightforward to calculate the structure factor 
$S(q)$ for the fluctuations of $\phi$ 
on the basis of $f_{\rm tot}$ 
in Eq.(6.6). As a function of the wave number $q$, 
it takes  the same  functional form as in Eq.(3.14), 
while the coefficients in the polyelectrolyte case 
are much more complicated than those 
in the electrolyte  case. 
That is, the shift $-(g_1+g_2)^2n_0/2$ in 
Eq.(3.14) is replaced 
by its counterpart 
$\Delta r$ dependent on $n_i$ and $n_p$ 
(for which see our paper\cite{Onuki-Okamoto}).  
In the following expressions (Eqs.(6.14)-(6.17)), 
$n_i$,  $n_p$, and $\phi$  represent  the average 
quantities. The Debye wave number of polyelectrolytes 
is given by \cite{Joanny}
\be 
\kappa^2= 4\pi \ell_B 
\bigg[(1-\zeta)n_p+\sum_{i} Z_i^2n_i\bigg],
\en 
which contains  the contribution from the  
(monovalent) ionized  
monomers ($\propto n_p$). 
The asymmetry parameter $\gamma_{\rm p}$ in Eq.(3.14) 
is of the form, 
\be 
\gamma_{\rm p}=({4\pi\ell_B}k_BT/{C})^{1/2}{A}/{\kappa^2}, 
\en 
where $C$ is given by  Eq.(6.5) and  
\be 
A= n_p/{\phi}  - (1-\zeta)g_pn_p+\sum_{i}Z_ig_i n_i .
\en 
In this definition, $\gamma_{\rm p}$ can  be negative 
depending on the  terms in $A$.
Mesophase formation can appear for  
 $|\gamma_{\rm p}|>1$ with increasing $\chi$.

The parameter $\gamma_{\rm p}$ is determined by  the 
  ratios among the charge densities and is  
  nonvanishing even in the dilute limit 
  of the charge densities.  
In particular, if  $n_c=n_p$ and $n_1=n_2$ 
in the monovalent case,  
$\gamma_{\rm p}$ is simplified as 
\be 
\gamma_{\rm p} =
\frac{{\phi}^{-1}-(1-\zeta)g_p+g_c+(g_1-g_2)R}{(4\pi\ell_BC/k_BT)^{1/2}
({2-\zeta+2R})}, 
\en 
where the counterions and 
the added cations are different. 
The  $R\equiv n_1/n_p$ is the  ratio 
between the salt density and the that of 
ionized monomers and Eq.(6.5) gives 
$(4\pi\ell_BC/k_BT)^{1/2}= 
[\pi \ell_B/3a\phi(1-\phi)]^{1/2}$. 

Some consequences follow from  Eq.(6.17). (i) 
With enriching a  salt  we eventually have 
  $R\gg |g_i|$; then,  the above formula tends to Eq.(3.15), 
  which is  
applicable for neutral polymer solutions (and 
low-molecular-weight 
binary mixtures for $N=1$) with salt. 
(ii) 
Without the solvation or for $g_i=0$,  
the above $S(q)$ tends to 
the previous  expressions  for 
polyelectrolytes \cite{Lu1,Lu2,Joanny},  
where $\gamma_{\rm p}$  decreases with increasing $R$. 
In accord with this,  
Braun et al. \cite{Candau} observed 
a mesophase    at low  salt contents and 
  macrophase  separation at high salt contents. 
%Thus more experiments are desirable to detect it 
%in  polyelectrolytes with water-like solvent. 
(iii) In our theory, neutral polymers 
in  a polar solvent can 
exhibit a mesophase  
for  large $|g_1-g_2|$.

 Hakim {\it et al}. \cite{Hakim,Hakim1} found  
a broad  peak  at an intermediate 
wave number in the scattering amplitude 
in (neutral)  polyethylene-oxide (PEO) 
in  methanol and in  acetonitrile 
by adding a small amount of  salt KI. They ascribed   
the origin of the peak  to 
binding of K$^+$  to PEO chains. 
%Remarkably, the peak 
% disappeared if the solvent was water, 
%which indicates  sensitive 
% dependence of the molecular interactions 
%on  the solvent species.  
Here more experiments are informative. 
An experiment by 
Sadakane {\it et al.} \cite{Sadakane} 
suggests that use of an antagonistic salt would  yield   
 mesophases more easily.

{\bf {6.5 Interface profiles without salt}}. 
We suppose coexistence of 
 two salt-free phases ($n_1=n_2=0$),  separated by a planar 
interface.   Even without salt, the interface profiles are  
extremely varied,  sensitively depending   
on the molecular interaction parameters, 
$\Delta_0$, $g_p$, and $g_c$. 
If a salt is added, they furthermore  depend on $g_1$, $g_2$, 
and the salt amount.  
The quantities with the subscript $\alpha$ 
($\beta$) denote 
the bulk values in the polymer-rich (solvent-rich)  phase  
attained as  $z \rightarrow -\infty$ 
(as  $z \rightarrow \infty$). The  ratio of the 
bulk counterion densities 
is given by 
%written as  $n_{c\alpha}= 
%n_c(-\infty)$ and $n_{c\beta}= n_c(\infty)$. 
%Here we obtain $n_{c\alpha}$ or  $n_{c\beta}$ from 
% Eq.(2.29) and  
\be 
\frac{n_{c\alpha}}{
n_{c\beta}}= \frac{\phi_\alpha\zeta_\alpha}{\phi_\beta\zeta_\beta}=
\exp\bigg[-g_c\Delta\phi-\frac{e\Delta \Phi}{k_BT}\bigg].
\en  
The Galvani potential difference 
$\Delta\Phi=\Phi_\alpha-\Phi_\beta$ 
is expressed in terms of 
$Q(\phi)$ in Eq.(6.13)  as    
\be 
\frac{
e\Delta \Phi }{k_BT} = g_p\Delta\phi+\ln \bigg[
\frac{\sqrt{Q(\phi_\beta)+1}-1}{\sqrt{Q(\phi_\alpha)+1}-1}\bigg],
%\frac{n_{c\beta}}{n_{c\alpha}}
%&=&\frac{\phi_\beta(\sqrt{Q_\alpha+1}+1)}{\phi_\alpha(\sqrt{Q_\beta+1}+1)}.
\en 
If $Q(\phi_\alpha)\gg 1$ and $Q(\phi_\beta) \gg 1$ 
(or $\zeta_\alpha\ll 1$ and $\zeta_\beta\ll 1$), we obtain 
$e\Delta \Phi/k_BT 
 \cong (g_p-g_c)\Delta\phi/2+\ln(\phi_\beta/\phi_\alpha)$.

%fig12
\begin{figure}[htbp]
\vspace{-35mm}
\begin{center}
 \includegraphics[scale=0.5, bb= 0 0 652 812]{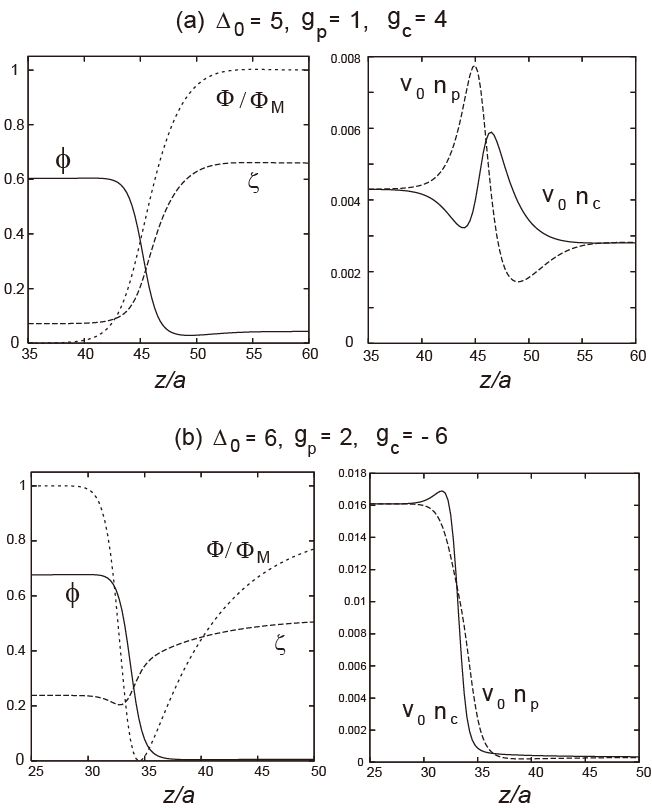} 
\caption{Interface profiles in the salt-free case 
for  (a) $\Delta_0=5$, $g_p=1$, and $g_c=4$ (top)  
and for (b) $\Delta_0=8$, $g_p=2$, and $g_c=-6$ (bottom). 
Polymer volume fraction $\phi(z)$, 
normalized potential $\Phi(z)/\Phi_M$, 
and degree of ionization $\zeta(z)$ (left), and 
normalized charge densities $v_0n_c(z)$ and  $v_0n_p(z)$ 
(right). The other parameters 
are common as  $\chi=1$,  $N=20$, $\nu_M=0.1$, 
$\ve_1=-0.9\ve_0$, and $\ell_B=8a/\pi$. 
Here $\Phi(z)$ is measured from its  
minimum,  and $\Phi_M (=2.67k_BT/e$ in (a) and $6.86k_BT/e$ in (b)) is 
the difference of its maximum and  minimum. 
$[$From: A. Onuki and R. Okamoto,  
J. Phys. Chem. B, {\bf 113}, 3988 (2009).$]$}
\end{center}
\end{figure}

%13
\begin{figure}[htbp]
\vspace{-30mm}
\begin{center}
 \includegraphics[scale=0.4,bb=0 0 699 890]{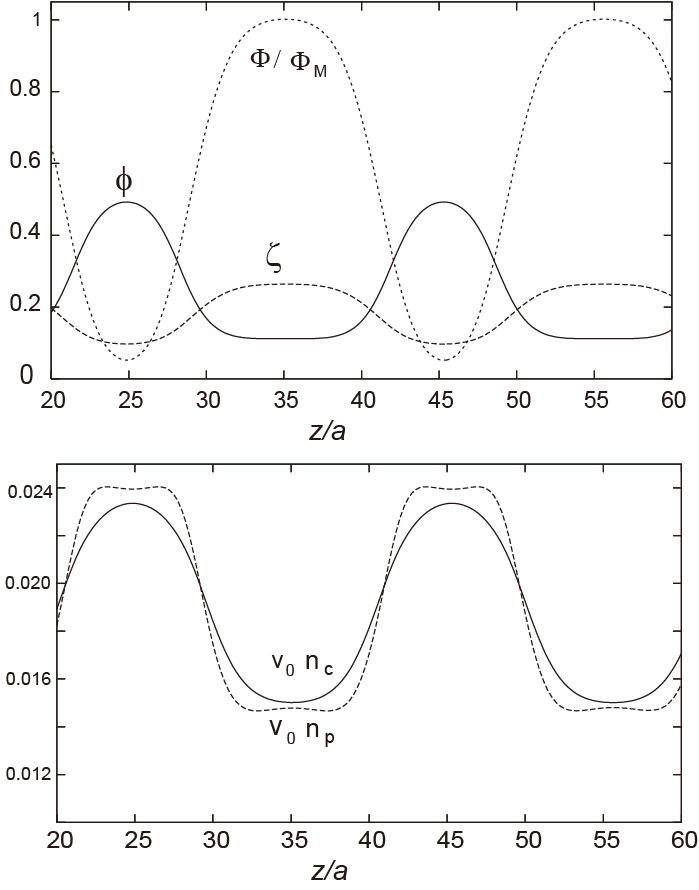} 
\caption{Periodic profiles in a salt-free 
mesophase  with   $\nu_M=0.5$,   $\Delta_0=5$, 
$g_p=1$, and $g_c=4$. 
  Top: $\phi(z)$. 
$\Phi(z)/\Phi_M$ with $\Phi_M=0.86T/e$, 
and  $\zeta(z)$.  Bottom: 
$v_0n_c(z)$ and  $v_0n_p(z)$. 
$[$From: A. Onuki and R. Okamoto,  
J. Phys. Chem. B, {\bf 113}, 3988 (2009).$]$ 
}
\end{center}
\end{figure}

We give  numerical results 
of one-dimensional profiles in equilibrium, 
where we  set $\chi=1$, $N=20$,   
$\ve_1=-0.9\ve_0$, and $\ell_B=e^2/\ve_0T= 
8a/\pi$.   The  dielectric constant 
of the solvent is 10 times larger than 
that of the polymer. The space will
 be measured in units of the 
molecular size $a=v_0^{1/3}$. 
In Fig.12, we show salt-free interface profiles  for 
(a) $\Delta_0=5$, $g_p=1$, and $g_c=4$   
and (b) $\Delta_0=8$, $g_p=2$, and $g_c=-6$.  
 In  the $\alpha$ and  $\beta$ regions, 
 the degree of ionization $\zeta$ 
is $0.071$ and $0.65$ in (a) 
and is $0.24$ and $0.51$ in (b), respectively. 
%For example, if $a=3{\rm \AA}$, 
%the counterion density  at  100 mM  is 
%equal to $n_c= 6\times 10^{19}$cm$^{-3}$ 
%and the normalized density $v_0n$ 
%is equal to $1.6\times 10^{-3}$. 
The normalized potential drop 
$e(\Phi_\alpha-\Phi_\beta)/k_BT$ 
is $-2.67$ in (a) and $0.099$ in (b). 
Interestingly, in (b), 
$\Phi(z)$ exhibits a deep minimum 
at the interface position.  
We can see appearance of the charge density 
$n_c-n_p$ around the interface, resulting in an electric 
double layer. The counterion density is 
shifted to the $\beta$ region in (a) 
because of positive $g_c$ 
and  to the $\alpha$ region in (b) 
because of negative $g_c$.  
The parameter $\gamma_{\rm p}$ 
in Eq.(6.16) is $0.75$ in (a) and $0.20$ in (b) 
in the $\alpha$ region, ensuring the stability 
of the $\alpha$ region.  

The surface tension $\sigma$ 
is again expressed as in Eq.(3.26) 
with  the negative 
electrostatic contribution. 
It  is calculated as 
$\sigma=0.0175k_BT/a^2$ in (a) and as $0.0556k_BT/a^2$ in (b), 
while we obtain    $\sigma = 0.050k_BT/a^2$ 
without ions at the same $\chi=1$. 
 In (a)  $\sigma$ 
is largely decreased because 
the electrostatic term $\sigma_{\rm e}$ in Eq.(3.27) 
is increased due to the formation of a large 
electric double layer. In (b), on the contrary,  it is increased by 
$10\%$ due to depletion of the charged particles  
from the interface \cite{OnukiPRE}.

We mention calculations of the interface profiles 
in weakly charged polyelectrolytes in a poor solvent 
using  self-consistent field theory 
\cite{Shi,Taniguchi}. 
In these papers, however, the solvation 
interaction was neglected.

{\bf {6.6 Periodic states  without salt}}. 
With varying the temperature (or $\chi$), 
the average composition $\av{\phi}$, 
the amount of salt, 
there can emerge a number of mesophases 
sensitively depending on the various 
molecular parameters ($g_i$, $\Delta_0$, and $\nu_M$). 
In Fig.13, we show an example of a one-dimensional 
periodic state without salt. Here   $\nu_M$ is 
set equal to  $0.5$ and the charge 
densities are much  increased.  In this case, 
the  degree of segregation 
and the charge heterogeneities 
are much   milder than in the cases in  Fig.12.

\section{Summary and  remarks} 

In this review, 
we  have tried to demonstrate  
the crucial role of the selective 
%hydrophilic and hydrophobic  
solvation of a solute in phase transitions of  various soft 
materials. We have used   coarse-grained 
approaches to investigate mesoscopic solvation 
effects. Selective solvation 
 should be relevant in  understanding a wide range  
of mysterious phenomena in water. 
Particularly remarkable  in polar binary mixtures are 
mesophase formation induced by  an antagonistic 
salt  and precipitation induced by a  one-sided 
solute (a salt composed of 
hydrophilic cations and anions  and a  neutral 
hydrophobic solute).  
Regarding the first problem, our theory 
is still insufficient and cannot 
well explain  the complicated  phase behavior 
disclosed by  the experiments 
\cite{Sadakane,Seto,SadakanePRL}. 
To treat  the second 
problem, we have started with the  free energy density 
$f_{\rm tot}$ in Eq.(5.1), which looks  rather 
obvious but yields highly nontrivial 
results  for large $g$. Systematic experiments 
are now possible. In particular, 
this precipitation takes place 
on  colloid surfaces 
as a prewetting phase transition  
near the precipitation curve $\chi=\chi_{\rm p}$ 
as in Fig.10 \cite{Okamoto}.

Though still preliminary, 
we have also treated an 
ionic surfactant system, where added in water-oil 
are  cationic surfactant, 
anionic counterions, 
and   ions from a salt. In this case, 
 we have introduced  the amphiphilic interaction 
 as well as  the solvation 
interaction to study 
the interface adsorption.   For ionic surfactants, 
the Gibbs formula \cite{Safran,Gibbs} 
for the surface tension is insufficient, because it 
neglects the electrostatic interaction.

In polyelectrolytes,  the charge distributions  
 are extremely complex 
around interfaces and in mesophases, 
sensitively depending on the 
molecular interaction   
and the dissociation process. 
Our continuum theory  takes  into account 
 these effects  in the  simplest  manner, 
though our results are still fragmentary. 
Salt effects in polyelectrolytes 
should also be further studied, 
on which  some discussions  can be 
found in our previous paper \cite{Onuki-Okamoto}.  
In the future, we should examine phase separation processes in 
polyelectrolytes, where the composition, 
the ion densities, and the degree of ionization 
are highly inhomogeneous. 
In experiments, large scale heterogeneities 
have been observed to be pinned 
in space and time \cite{Ise,Amis}, giving rise to 
enhanced scattering at small wave numbers.

As discussed in Sec.1,   there can be 
phase separation induced by 
selective hydrogen bonding. In particular, 
the effect of moisture uptake is 
dramatic in  PS-PVME \cite{Hashimoto}, where 
scattering experiments controlling the water content 
are desirable. To investigate 
 such polymer blends theoretically,  we may  
use the form in Eq.(5.1) with $n$ being the 
water density and $f(\phi)$ being 
the Flory-Huggins free energy 
for  polymer blends\cite{PG}. 
Similar problems should also be 
encountered  in block polymer systems containing  
ions or water. It is also known  that 
blends  of block polymer and homopolymer 
exhibit  complicated phase behavior 
for different interaction 
parameters $\chi_{ij}$ \cite{Shi1}.

We mention two  interesting effects  not discussed  
in this review. First, there can be an intriguing 
interplay between the  solvation and 
the hydrogen bonding in phase separation. 
For example, in some aqueous  mixtures, 
even  if they are  miscible at all $T$ 
at atmosphere pressure without salt, addition of  
a  small amount of a hydrophilic  salt 
gives  rise to reentrant phase separation behavior 
\cite{Kumar,cluster,Anisimov,Misawa}. 
On the other hand, Sadakane {\it et al.} 
observed a shrinkage of a closed-loop coexistence curve 
by adding an atagonistic salt or an ionic surfactant \cite{Seto}. 
Second, molecular polarization  of  polar  molecules or ions can 
give rise to a surface potential difference  on the 
molecular scale at an interface. See  such an example 
  for water-hexane \cite{Patel}. 
This effect is particularly noteworthy 
for hydronium ions in acid solutions 
\cite{Levinhyd}. 

We  will  report on  the 
wetting transition on  charged walls, rods, 
and  colloids and the solvation-induced colloid 
interaction. These effects are  
 much influenced  by the ion-induced 
precipitation discussed in Sec.5. 
In these  problems, 
first-order prewetting transitions  occur from 
weak-to strong  ionization and adsorption, 
as discussed in our paper on charged rods  
\cite{Oka}.  We will also report 
that  a small amount of 
a hydrophobic solute can produce small 
bubbles in water even outside the coexistence curve, 
on which there have been a large number of experiments.

\vspace{2mm}
\noindent{\bf Acknowledgments}\\
\noindent 
This work was supported by 
KAKENHI (Grant-in-Aid for Scientific Research)
on Priority Area  Soft Matter Physics from
the Ministry of Education, Culture, 
Sports, Science and Technology of Japan. 
Thanks are due to informative discussions 
with   M. Anisimov, 
K. Sadakane, H. Seto, T.Kanaya, K. Nishida, 
 T. Osakai, T. Hashimoto, and F. Tanaka.

\vspace{2mm} 
{\bf Appendix A: Statistical theory 
of selective solvation at small water composition}\\
\setcounter{equation}{0}
\renewcommand{\theequation}{A\arabic{equation}}

We present a simple statistical theory  of 
 binding  of polar molecules 
 to hydrophilic ions  due to the ion-dipole 
 interaction in a water-oil mixture \cite{Oka} 
 when the water volume fraction $\phi$  is  small.  
We assume no macroscopic  inhomogeneity 
and do not treat   
the large-scale electrostatic interaction.  
Similar arguments were given for hydrogen boning between 
water and  polymer \cite{hydrogen1,hydrogen2}.

Our  system has  a volume $V$ and contains  
$N_{\rm w}$ water molecules. Using  the water 
molecular volume $v_0$, 
we have 
\be 
\phi= N_{\rm w}v_0 /V= N_{\rm w}/N_0,
\en 
where $N_0=V/v_0$. 
We fix $N_{\rm w}$ or $\phi$ in the following. 
The total ion numbers are denoted by $N_{{\rm I}i}=V n_i$, 
where $i=1$ for the cations and $i= 2$ 
for the anions with  $n_i$ being  the average densities.  
The ionic  volumes are assumed to be small 
and their  volume fractions are neglected. 
Then the  oil volume fraction is given by 
$1-\phi$. 
 Each solvation shell  
consists  of $\nu$ water molecules 
with $ \nu=1,\cdots, Z_i$, where  $Z_i$ 
is  the maximum water number in a shell. 
%The   free energy decreases by $k_B Tw_{i\nu}$ 
%upon a cluster formation.  
%If  $w_{i\nu} \gg 1$, the binding   
%can be significant  even for small  $\phi$. 
%From Eq.(2.3) 
%we estimate $w_{i\nu}/\nu \sim Z_i^2
%e^2\ve_1/k_BTR_i\ve^2$, 
%where $\ve$ is close to the oil dielectric 
%constant at small $\phi$. 

Let  the number of the $\nu$-clusters 
composed of $\nu$ water molecules  around an ion be 
$\gamma_{i\nu} N_{\rm w}$.   
The total number of the solvated    ions  
is then $\gamma_i N_{{\rm I}i}$ with 
\be 
\gamma_i=\sum_{\nu} \gamma_{i\nu}<1, 
\en  
where $1\le \nu \le Z_i$. 
The number of the bound water molecules  
in  the $\nu$-clusters  is 
$\nu \gamma_{i\nu}N_{{\rm I}i}$. 
The fraction  of the unbound water molecules  $\phi_{{\rm f}}$ satisfies   
\be 
 \phi_{{\rm f}} +v_0n_i \sum_{i, \nu} 
  \nu \gamma_{i\nu} =\phi.
\en

We construct the free energy of  the 
total system $F_{\rm tot}$ 
for each given set of $\gamma_{i\nu}$. 
In terms of the oil density 
$n_{\rm oil}=v_0^{-1}(1-\phi)$, 
the   unbound water density 
$n_{\rm wf}=v_0^{-1}\phi_{{\rm f}}$,  
the unbound ion densities 
$n_{i{\rm f}}= n_i(1-\gamma_i)$, and the cluster densities 
$n_{i\nu}= n_i \gamma_{i\nu}$,
we obtain 
\bea 
&&\frac{F_{\rm tot}}{Vk_BT} 
= n_{\rm oil} [\ln(n_{\rm oil}\lambda_{\rm oil}^3) -1] +   
 n_{\rm wf} [\ln(n_{\rm wf}\lambda_{\rm w}^3) -1]  \nonumber\\
&& +
  \sum_i [n_{if } \ln (n_{i{\rm f}}\lambda_i^3]-n_i ]
+ \sum_{i, \nu} n_{i\nu}
[\ln  (n_{i\nu}\lambda_{i\nu}^3) - w^0_{i\nu}] \nonumber\\
&&+   
 \chi v_0 n_{\rm oil}n_{\rm wf} + \sum_{i, \nu} \chi_{i\nu} v_0 n_{i\nu}n_{\rm oil},
\ena 
where $\lambda_{\rm oil}$, $\lambda_{\rm w}$, $\lambda_{i}$, 
and $\lambda_{i\nu}$ are the thermal 
de Broglie wavelengths, $k_BTw^0_{i\nu}$ are  the "bare" 
binding free energies, 
and $\chi$ is the interaction parameter 
between the unbound water  and the oil. 
We asssume short-range 
interactions among the clusters and the oil 
characterized by the interaction 
parameters $\chi_{i\nu}$ to obtain the last term. 
At   small $\phi$, 
 the interactions among the clusters 
and the unbound water are neglected.  That is, 
we neglect the contributions of order $\phi^2$. 
We then calculate 
the solvation contribution 
 $F_{\rm sol}\equiv F_{\rm tot}-F_0$, 
where $F_0$ is the free energy without binding 
($\gamma_{i\nu}=0$). Some calculations give 
\bea
\frac{F_{{\rm sol}}}{N_0 k_B T }&=&
\phi_{{\rm f}} (\ln\phi_{{\rm f}}-1) +   
  \sum_i n_{ i } 
  (1-\gamma_i)\ln (1-\gamma_i)
\nonumber\\
&&\hspace{-1cm}  +   \sum_{i, \nu} n_{i } 
 \gamma_{i\nu}(\ln  \gamma_{i\nu} -w_{i\nu}) 
 - \phi(\ln\phi-1),
\ena  
where $k_BT w_{i\nu}$ are the ''renormalized''  
binding free energies written as 
\be 
w_{i\nu}=w^0_{i\nu}+ 
3\ln(\lambda_i\lambda_{\rm w}^\nu/\lambda_{i\nu}) 
+\nu \chi- \chi_{i\nu} .
\en  
The fractions $\gamma_{i\nu}$ are determined by 
minimization of  $F_{\rm sol}$ with respect to 
$\gamma_{i\nu}$ under Eqs.(A2) and (A3) as   
\bea 
{\gamma_{i\nu}} &=& 
({1-\gamma_i})\phi_{{\rm f}}^\nu e^{w_{i\nu}}, \\  
\gamma_i &=&  1-1/(1+\sum_{\nu} \phi_{{\rm f}}^\nu e^{w_{i\nu}}).
\ena
Substitution of Eqs.(A7) and (A8) into Eq.(A5) yields 
\be
\frac{F_{\rm sol}}{k_BTN_0}= 
\phi\ln\frac{\phi_{{\rm f}}}{\phi}  + \phi-\phi_{{\rm f}}  
+ \sum_i v_0n_i 
 \ln(1-\gamma_i).
\en

First, we assume the dilute limit of ions $N_I\ll N_{\rm w}$, where  
we have $\phi-\phi_{{\rm f}}  \ll \phi$.  
In the right hand side of Eq.(A9),  the sum of 
the first three terms  
 becomes  $- (\phi-\phi_{{\rm f}})^2/2\phi$ 
and is negligible.      We write  ${F_{\rm sol}}$ 
 as the sum $ V\sum_in_i\mu_{{\rm sol}}^i (\phi) $ 
 and use  Eq.(A8) to obtain 
 the solvation chemical potentials of the form,     
\be 
\mu_{{\rm sol}}^i (\phi) 
=  -  k_BT \ln (1+\sum_{\nu}  \phi^\nu e^{w_{i\nu}}). 
\en 
Let the maximum  of 
$ k_BT w_{i\nu}/\nu$ 
(per molecule  for various $\nu$) 
be  $\epsilon_{{\rm b}i}$ for each $i$ 
(see Eq.(2.3)). Then 
we obtain  the expression (2.2) for 
the crossover  volume fraction 
$\phi_{\rm sol}^i$.  
 For $\phi > \phi_{\rm sol}^i$, 
 $\gamma_i$ approaches unity.

Second,  we consider the dilute limit 
of water,  $\phi \ll \phi_{\rm sol}^i$ ($i=1,2$),  
where  the cluster fractions $\nu_i$ are small and 
the dimers with $\nu=1$ 
are dominant as indicated 
in the experiment  \cite{proton}. Neglecting the contributions 
from the clusters with $\nu \ge 2$,  we obtain 
\be 
\gamma_i\cong \phi_{{\rm f}} e^{w_{i1}},\quad 
\phi_{{\rm f}}\cong \phi/(1+ S),
\en   
where $S$ is the parameter defined as 
\be 
S=   v_0(n_1e^{w_{11}}+n_2e^{w_{21}}) . 
\en 
The solvation free energy behaves as   
\be   
F_{\rm sol}/N_0 k_BT \cong - 
\phi \ln (1+ S).   
\en   
For $S \ll 1$, 
we find $\mu_{{\rm sol}}^i (\phi)\cong -k_BT \phi e^{w_{1i}}$. 
However, if $S\gs 1$,  
the solvation chemical potential are not well defined.

\vspace{2mm} 
{\bf Appendix B:  Ions at liquid-liquid interface }\\
\setcounter{equation}{0}
\renewcommand{\theequation}{B\arabic{equation}}
 
In electrochemistry,  attention 
has been paid  to the ion distribution 
 and the electric potential difference 
 across a liquid-liquid  interface \cite{Ham,Hung}. 
 (In the vicinity of  an air-water interface,  
  virtually no ions are  
  present in the bulk air region 
 \cite{Onsager,Levin-Flores}.)
%With formation of a well-defined 
%solvation shell, the typical 
%magnitude of $\mu_{\rm sol}^i$ should 
%much exceed the thermal energy $k_BT$. 
Let us suppose  two species of ions ($i=1, 2$) 
with charges $Z_1e$ and $Z_2e$ ($Z_1>0$, $Z_2<0$). 
At  low ion densities, 
the  total ion chemical potentials $\mu_i$ in a  mixture solvent  
are expressed as  
\be 
\mu_i= k_BT \ln (n_i\lambda_i^3) 
+ Z_ie\Phi+ \mu_{\rm sol}^i (\phi),
\en 
where $\lambda_i$ is the thermal de Broglie length 
(but is an irrelevant constant in the isothermal condition)  
and $\Phi$  is the local electric potential. 
This quantity is a constant in equilibrium. 
For neutral hydrophobic particles 
the electrostatic term is nonexistent, 
so we have  Eq.(2.5).

We consider  a liquid-liquid  interface  between  
a  polar (water-rich) 
phase $\alpha$ and a less polar (oil-rich) 
phase $\beta$ with bulk compositions $\phi_\alpha$ 
and $\phi_\beta$ with $\phi_\alpha>\phi_\beta$. 
The bulk ion densities far from the interface are written as 
$n_{i\alpha}$ in phase $\alpha$ and 
$n_{i\beta}$ in phase $\beta$.  From   the 
charge neutrality condition in  the bulk regions, 
 we  require 
\be 
Z_1n_{1\alpha}+Z_2n_{2\alpha}=0, \quad 
   Z_1n_{1\beta}+Z_2n_{2\beta}=0. 
\en    
The potential    $\Phi$ tends to 
 constants  
$\Phi_\alpha$ and $\Phi_\beta$ in the  bulk 
 two phases, yielding   a  Galvani potential difference, 
$ 
 \Delta\Phi=\Phi_\alpha-\Phi_\beta. 
$ 
Here  $\Phi$ approaches 
its limits  on the  scale of the Debye 
screening lengths,  $\kappa_\alpha^{-1}$ and 
$\kappa_\beta^{-1}$,  away from the interface, 
so we assume that 
the system  extends longer than 
$\kappa_\alpha^{-1}$ in phase $\alpha$ 
and $ \kappa_\beta^{-1}$ in phase $\beta$. 
Here we neglect 
molecular polarization of solvent molecules 
and surfactant molecules 
at an interface (see 
comments in the summary section).

The solvation chemical potentials 
$\mu_{\rm sol}^i(\phi)$ also take different values 
in the two phases due to 
their  composition dependence. So we define 
  the differences   
$\Delta\mu_{\alpha\beta}^{i}
$ as in Eq.(2.4). 
The continuity of $\mu_i$ across the interface  gives  
\be 
k_BT \ln (n_{i\alpha}/n_{i\beta})+ Z_i e\Delta\Phi 
- \Delta\mu_{\alpha\beta}^{i}=0,  
\en 
where $i=1,2$. From Eqs.(B2) and (B3), 
the Galvani potential difference 
 $\Delta\Phi$ is expressed as 
 \cite{Hung,OnukiPRE}
\be
\Delta \Phi=[{\Delta\mu_{\alpha\beta}^1 
-\Delta\mu_{\alpha\beta}^{2}}]/{e(Z_1+|Z_2|)}. 
\en 
Similar potential differences also appear 
at liquid-solid interfaces (electrodes) \cite{Ham}. 
The ion densities in the bulk two phases (in the dilute limit) 
are simply related by  
\be 
 \frac{n_{1\beta}}{n_{1\alpha}}
= \frac{n_{2\beta}}{n_{2\alpha}} = 
\exp\bigg[- \frac{|Z_2| {\Delta\mu_{\alpha\beta}^{1}} +
Z_1 \Delta\mu_{\alpha\beta}^{2} }{(Z_1+|Z_2|)k_BT}\bigg] .  
\en 
However, if  three 
ion species are present,  the ion partitioning 
between two phases is much  more complicated 
\cite{OnukiJCP}.

%\section*{References}

\end{document}